\pdfoutput=1
\documentclass{ectj}
\usepackage{amsfonts,amssymb,graphics,epsfig,verbatim,bm,latexsym,amsmath,url,amsbsy}
\graphicspath{ {./img/} }
\usepackage{bbm, listings}
\usepackage{tikz-cd}

\newtheorem{assumption}{Assumption}
\newtheorem{proposition}{Proposition}
\newtheorem{corollary}{Corollary}
\newtheorem{lemma}{Lemma}
\newtheorem{example}{Example}
\newtheorem{remark}{Remark}

\newtheorem{definition}{Definition}
\newtheorem{notation}{Notation}

\renewcommand{\thesection}{\arabic{section}}
\renewcommand{\theequation}{\arabic{section}.\arabic{equation}}

\newcommand\posscite[1]{\citeauthor{#1}'s (\citeyear{#1})}

\received{February 2017}
\accepted{November 2017}
\volume{20}
  
\title[Nonparametric Bounds on Treatment Effects with Imperfect Instruments]{Nonparametric Bounds on Treatment Effects with Imperfect Instruments}

\author[Kyunghoon Ban and D\'esir\'e K\'edagni]{Kyunghoon Ban$^{\dagger}$ and
                        D\'esir\'e K\'edagni$^{\dagger}$}

\address{$^{\dagger}$Department of Economics, Iowa State University,\\
                    260 Heady Hall, 518 Farm House Lane, Ames, IA, 50011, USA.}

\email{khban@iastate.edu, dkedagni@iastate.edu}

\def\AmSTeX{$\cal A$\kern-.1667em\lower.5ex\hbox{$\cal M$}\kern-.125em
            $\cal S$-\TeX}
\def\BibTeX{{\rm B\kern-.05em{\sc i\kern-.025em b}\kern-.08em
            T\kern-.1667em\lower.7ex\hbox{E}\kern-.125emX}}

\begin{document}
\nonstopmode

  \begin{abstract}
   This paper extends the identification results in \cite{Nevo2012} to nonparametric models. We derive nonparametric bounds on the average treatment effect when an imperfect instrument is available. As in \cite{Nevo2012}, we assume that the correlation between the imperfect instrument and the unobserved latent variables has the same sign as the correlation between the endogenous variable and the latent variables. We show that the monotone treatment selection and monotone instrumental variable restrictions, introduced by \cite{Manski2000, Manski2009}, jointly imply this assumption.
Moreover, we show how the monotone treatment response assumption can help tighten the bounds. The identified set can be written in the form of intersection bounds, which is more conducive to inference. We illustrate our methodology using the National Longitudinal Survey of Young Men data to estimate returns to schooling.


  \end{abstract}

\section{Introduction}

   The use of an instrumental variable (IV) is a popular solution to deal with endogeneity in social sciences. However, this approach may yield misleading conclusions when the instrument is invalid. A valid instrument must be uncorrelated with the unobservables in the model.\footnote{A stronger version of this condition is that the instrument is statistically (or mean) independent of the unobservables.} This requirement is often difficult to justify, and may call into question empirical findings. For this reason, \cite{Nevo2012} derived bounds on the parameters of interest (e.g., the average treatment effect) in parametric models under weaker conditions. They first assume that the sign of correlation between the imperfect IV (IIV)\footnote{An instrument that is potentially correlated with the unobservables.} and the unobserved latent variables is the same as that of the correlation between the endogenous variable and the latent variables. Second, they add the assumption that the correlation between the IIV and the latent variables is less than the correlation between the endogenous variable and the latent variables to tighten the bounds on the parameters of interest.

In this paper, we derive nonparametric bounds on the average treatment effect with an imperfect IV under the above assumptions when the outcome variable has bounded support. We introduce the concept of binarized MTS-MIV, which is implied by the monotone treatment selection (MTS) and monotone IV (MIV) assumptions developed by \cite{Manski2000, Manski2009}. We show that the correlation between a binarized MTS-MIV and the unobserved latent variables has the same sign as the correlation between the endogenous variable and the latent variables. Hence, we link the \cite{Nevo2012} same direction of correlation assumption to the \cite{Manski2000} monotone treatment selection and monotone IV assumptions.  
We believe this result is new in the literature. Furthermore, we show how additional restrictions such as the less endogenous instrument, and the monotone treatment response can help tighten the bounds. As in \cite{Nevo2012}, the bounds take the form of intersection bounds and can be implemented using the inferential methods developed by \cite{CLR2013} or \cite{AS2013}. We illustrate our methodology using the National Longitudinal Survey of Young Men (NLSYM) data to estimate returns to schooling.

There is an increasing interest in the identification of causal effects with imperfect instrumental variables. Recently, \cite{Masten2020} developed a methodology that allows researchers to consider continuous relaxations of the IV models when they are refuted by the data. Their approach is data driven as it exploits the extent of falsification of the model to construct the identified set for the parameter of interest. Although their method helps salvage some invalid IVs, their identifying assumptions may seem difficult to interpret. Our approach, as well as that of \cite{Nevo2012} and \cite{Manski2000, Manski2009}, is not data driven and has clearer interpretations of the identifying assumptions. This paper also relates to the work of \cite{Kedagni2018b}, who consider weaker version of the mean independence assumption for the IV model. They derived nonparametric bounds on the average treatment effect under unconditional moment restrictions for the IV. Several other papers have also studied identification of model parameters when the IV is invalid using a framework different from ours; see \cite{Hotz1997}, \cite{Conley2012}, among others.

The remainder of the paper is organized as follows. Section \ref{anaF} presents the model, the assumptions and their link with the literature. In Section \ref{ident}, we derive our main identification results. We discuss inference and implementation in Section \ref{inference}. Section \ref{empirical} presents an empirical illustration of our proposed methodology, while Section \ref{conclusion} concludes. Proofs and additional results are relegated to the appendix.

\section{Analytical Framework}\label{anaF}
Consider the following potential outcome model (POM)
\begin{eqnarray}\label{eq:pom}
Y=\sum_{d=1}^T Y_d\mathbbm{1}\left\{D=d\right\},
\end{eqnarray}
where $Y$ is the outcome variable taking values in $\mathcal Y \subset \mathbb R$, $D$ is a discrete endogenous treatment variable taking values in $\mathcal D=\{1,2,\ldots,T\}$, $Y_{d}$ is the potential outcome that would have been observed if the treatment $D$ had externally been set to $d$. Let $Z\in \mathcal Z \subseteq \mathbb R$ be an imperfect IV in the sense that it may be correlated with the potential outcome $Y_d$. In what follows, we assume that the random variable $Y_d$ is integrable, i.e., $\text{E}[Y_d]<\infty$. The objects of interest in this paper are the potential outcome means $\theta_d \equiv \text{E}[Y_d]$, for all $d \in \mathcal D$, and some treatment effects $ATE(d,d')\equiv \theta_d-\theta_{d'}$, for $d,d' \in \mathcal D$. We allow for heterogeneous treatment effects, so that $ATE(d,d')$ may vary across $(d,d')$. The methodology that we develop in this paper can also be used to identify other commonly used parameters of interest such as the average treatment effect on the treated $ATT(d,d')\equiv\text{E}[Y_d-Y_{d'}\vert D=d]$, and the average treatment effect on the untreated $ATU(d,d')\equiv\text{E}[Y_d-Y_{d'}\vert D=d']$. But, for the sake of clarity of the exposition, we focus our attention on the $ATE$. 

We observe a random sample of the vector $(Y,D,Z)$. For simplicity, we drop exogenous covariates from the analysis. 
For example, $Y$ could be earnings, $D$ years of schooling, and $Z$ parental education. In this example, $Y_d$ is the potential earnings for an individual with $d$ years of schooling. 
We now state our main identifying assumptions:
\begin{assumption}[Bounded support (BoS)]\label{BS}
\begin{eqnarray*}
Supp(Y_d\vert D\neq d)= Supp(Y_d \vert D = d) =\left[\underline{y}_d,\overline{y}_d\right]\ \text{ for each }\ d \in \mathcal{D}.
\end{eqnarray*}
\end{assumption}
Assumption BoS states that the support of the counterfactual outcome is the same as that of the factual. It is standard and similar to the usual bounded outcome assumption considered in \cite{Manski1990, Manski1994}, and many other papers. Like in \cite{Kedagni2018b}, it allows the support of the potential outcome $Y_d$ to vary across all treatment levels $d$.

\begin{assumption}[Same direction of correlation (SDC)]\label{SDC}
\begin{eqnarray*}
Cov\left(Y_d,D\right) Cov\left(Y_d,Z\right)\geq 0\  \text{ for each }\ d \in \mathcal{D}.
\end{eqnarray*}
\end{assumption}

Assumption SDC is equivalent to Assumption 3 in \cite{Nevo2012}.  It states that the correlation between the imperfect instrument $Z$ and the potential outcome $Y_d$ has weakly the same sign as the correlation between the endogenous treatment $D$ and the potential outcome. For example, it is documented that parental education is not a valid instrument; see \cite{Kedagni20}, \cite{Mourifie2020}, among many others. However, one could assume that parental education has the same sign of correlation with the potential earnings as does the individual's education. Note that if either the treatment $D$ or the instrument $Z$ is exogenous, this assumption holds. If Assumption BoS holds from the definition of the outcome variable (e.g., market share lies between 0 and 1),
Assumption SDC has a testable implication. Indeed, when BoS holds and the bounds derived in Proposition \ref{thm1} for $\theta_d$ under BoS and SDC are empty, then SDC is rejected.

Assumption SDC can be seen as a weaker version of the concepts of monotone IV (MIV: $\text{E}\left[Y_d \vert Z=z\right]$ is monotone in $z$ for all $d$) and monotone treatment selection (MTS: $\text{E}\left[Y_d \vert D=\ell\right]$ is monotone in $\ell$ for all $d$) developed by \cite{Manski2000, Manski2009}. To show this result, we introduce the concept of binarized MTS-MIV that is intermediate between SDC and MTS-MIV. 
We use the following notation.
\begin{notation}
Denote $g_d^+(j)=\text{E}[Y_d|D\geq j]$, $g_d^-(j)=\text{E}[Y_d|D < j]$, $h_d^+(z)=\text{E}[Y_d|Z\geq z]$, $h_d^-(z)=\text{E}[Y_d|Z < z]$. 
$\rho_{UV}$ denotes the coefficient of correlation between two random variables $U$ and $V$.
\end{notation}
\begin{definition}
The variable $Z$ is a \textit{binarized MTS-MIV} for $D$ if for each $d \in \mathcal D$,
\begin{eqnarray}
\left(g_d^+(j)-g_d^-(j)\right)\left(h_d^+(z)-h_d^-(z)\right)\geq 0\ \text{ for all } j,\ z. \label{eq:BMTSMIV}
\end{eqnarray}
\end{definition}
In words, we say that $Z$ is a binarized MTS-MIV for $D$ if all binarized treatments $\mathbbm{1}\{D\geq j\}$ satisfy the MTS restriction, and all binarized instruments $\mathbbm{1}\{Z\geq z\}$ satisfy the MIV restriction.
\begin{remark}
\textnormal{
If $Z$ is a binarized MTS-MIV for $D$ then the functions $g_d^+$ and $g_d^-$ do not cross, nor do the functions $h_d^+$ and $h_d^-$ for all $d$; that is, either [$g_d^+(j) \geq g_d^-(j)$ for all $j$ and $h_d^+(z) \geq h_d^-(z)$ for all $z$] or [$g_d^+(j) \leq g_d^-(j)$ for all $j$ and $h_d^+(z) \leq h_d^-(z)$ for all $z$]. Moreover, if $g_d^+ \geq g_d^-$ for some $d$ then $h_d^+ \geq h_d^-$, and vice versa.
}
\end{remark}
Lemma \ref{lem1} shows that MTS-MIV is a sufficient condition for binarized MTS-MIV, while Lemma \ref{lem2} shows that binarized MTS-MIV is a sufficient condition for SDC.
\begin{lemma}\label{lem1}
MTS-MIV in the same direction for $D$ and $Z$ implies that $Z$ is a binarized MTS-MIV for $D$.
\end{lemma}

\begin{lemma}\label{lem2}
If $Z$ is a binarized MTS-MIV for $D$, then Assumption SDC holds.
\end{lemma}
\begin{remark}
\textnormal{
From Lemmas \ref{lem1} and \ref{lem2}, we conclude that MTS-MIV in the same direction implies Assumption SDC. Moreover, when both the treatment $D$ and the imperfect instrument $Z$ are binary, MTS-MIV in the same direction,  binarized MTS-MIV and Assumption SDC are equivalent. However, Example \ref{ex.0921} in the appendix shows a case where binarized MTS-MIV holds, but the joint MTS-MIV fails.} 

\textnormal{
If MTS and MIV hold in the opposite directions, then SDC will not hold. Instead, the ``opposite directions of correlation''  assumption $Cov(Y_d,D) Cov(Y_d,Z) \leq 0$ holds. The identification strategy developed in this paper can easily be adapted to this case. In general, if the directions of the MTS and MIV assumptions are unknown, SDC is not weaker than MTS-MIV.
}
\end{remark}

Another sufficient condition for binarized MTS-MIV is the joint positive quadrant dependence between the potential outcome $Y_d$ and the instrument $Z$, and between the potential outcome $Y_d$ and the treatment $D$. 
The concept of positive quadrant dependence has been considered by \cite{BSV2012} in a different framework. Two random variables $\varepsilon$ and $\nu$ are positive quadrant dependent (PQD) if 
\begin{eqnarray*}
\text{P}(\varepsilon \leq t_0 \vert \nu < t_1) \geq \text{P}(\varepsilon \leq t_0)\ \text{ for all }\ t_0, t_1.
\end{eqnarray*}
As pointed out by \cite{BSV2012}, the PQD assumption implies that
\begin{eqnarray*}
\text{P}(\varepsilon \leq t_0 \vert \nu < t_1) \geq \text{P}(\varepsilon \leq t_0 \vert \nu \geq t_1)\ \text{ for all }\ t_0, t_1.
\end{eqnarray*}
From this implication, we conclude that if $Y_d$ and $Z$ are PQD, then the distribution $Y_d$ conditional on $\{Z \geq z\}$ first-order stochastically dominates that of $Y_d$ conditional on $\{Z < z\}$ for all $z$. Therefore, $\text{E}[Y_d \vert Z \geq z] \geq \text{E}[Y_d \vert Z < z]$, i.e., $h_d^+(z)-h_d^-(z) \geq 0$ for all~$z$. Similarly, if $Y_d$ and $D$ are PQD, then $g_d^+(j)-g_d^-(j) \geq 0$ for all $j$. Hence, binarized MTS-MIV holds.

Note that joint positive quadrant dependence between $Y_d$ and $Z$,  and between $Y_d$ and $D$ does not imply joint MTS-MIV, and the converse does not hold either. However, joint positive regression dependence (PRD) between $Y_d$ and $Z$ (i.e., $\text{P}(Y_d > y \vert Z=z)$ is nondecreasing in $z$ for all $y$), and between $Y_d$ and $D$ implies both joint MTS-MIV, and joint positive quadrant dependence between $Y_d$ and $Z$,  and between $Y_d$ and $D$. See the proof in the appendix. Figure \ref{fig:imp} below summarizes the relationship between the different concepts.

\begin{figure}[!htbp]
        \centering
        \begin{tikzcd}[arrows=Rightarrow]
		\textrm{Joint PRD} \arrow[r] \arrow[d] & \textrm{Joint PQD} \arrow[d] \arrow[dr]\\
		\textrm{Joint MTS-MIV} \arrow[r] & \textrm{Binarized MTS-MIV} \arrow[r] & \textrm{SDC}
		\end{tikzcd}
        \caption{Implications Between Assumptions}\label{fig:imp}
\end{figure}
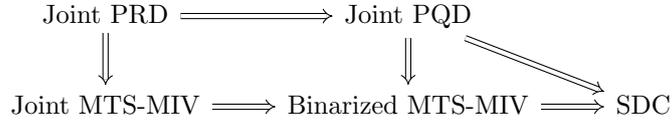

\begin{assumption}[Less endogenous instrument (LEI)]\label{LEI}
\begin{eqnarray*}
\mid\rho_{Y_d D}\mid \geq \mid \rho_{Y_d Z}\mid \text{ for each }\ d \in \mathcal D.
\end{eqnarray*}
\end{assumption}

Assumption LEI is the same as Assumption 4 in \cite{Nevo2012}, which they refer to as the ``instrument less endogenous than treatment'' assumption. It states that the imperfect instrument $Z$ is less correlated with the potential outcome than is the endogenous treatment $D$. In this paper, we use the shorthand ``less endogenous instrument'' to call this assumption. In the context of our empirical example, it is reasonable to assume that parental education is less correlated with the individual's potential wage than is the individual's own education.

\begin{assumption}[Monotone treatment response (MTR)]\label{MTR}
\begin{eqnarray*}
Y_d \geq Y_{d'}\ \text{ for all }\ d>d'. 
\end{eqnarray*}
\end{assumption}
Assumption MTR states that the potential outcome weakly increases with the level of the treatment. It was introduced by \cite{Manski1997}, and considered in \cite{Manski2000, Manski2009}, among many others. For instance, in the returns to schooling example, it implies that the wage that a worker earns weakly increases as a function of the worker's years of schooling. We show how this assumption can help tighten the bounds derived under Assumptions BoS, SDC and LEI.

Now that we have discussed the model and our identifying assumptions, we are going to present our main identification results.

\section{Identification results}\label{ident}
In this section, we derive under the different assumptions discussed in the previous section bounds on the potential outcome expectation $\theta_d$, for each $d \in \mathcal D$: $LB_d \leq \theta_d \leq UB_d$. Bounds on $ATE(d,d')$ are then obtained as: $LB_d-UB_{d'} \leq ATE(d,d') \leq UB_d-LB_{d'}$.

\setcounter{equation}{0}
\subsection{Identification under the same direction of correlation assumption}
Assumption SDC is equivalent to $\text{E}\left[Y_d\tilde{D}\right]\text{E}\left[Y_d\tilde{Z}\right] \geq 0$, where $\tilde{D}\equiv D-\text{E}[D]$ and $\tilde{Z}\equiv Z-\text{E}[Z]$, which in turn is equivalent to: either
\begin{eqnarray}
    \text{E}\left[Y_d\tilde{D}\right] \geq 0  \ \text{  and  } \ \text{E}\left[Y_d\tilde{Z}\right] \geq 0,
    \label{eq.SDC1}
\end{eqnarray}
 or 
\begin{eqnarray}
    \text{E}\left[Y_d\tilde{D}\right] \leq 0 \ \text{  and  } \ \text{E}\left[Y_d\tilde{Z}\right] \leq 0.
    \label{eq.SDC2}
\end{eqnarray}
We first derive bounds on the potential outcome mean $\theta_d$ using inequalities (\ref{eq.SDC1}). Similarly, we can derive the bounds implied by inequalities (\ref{eq.SDC2}). 

Inequality (\ref{eq.SDC1}) implies that, for all $(\lambda,\gamma)\in \mathbb R^2_+\setminus \left\{(0,0)\right\}$, we have
\begin{eqnarray*}
\text{E}\left[Y_d \left(\lambda\tilde{D}+\gamma\tilde{Z}\right)\right] \geq 0.
\end{eqnarray*} 
By factorizing $(\lambda + \gamma)$ in the above inequality, we have
\begin{eqnarray*}
    \text{E}\left[Y_d (\lambda + \gamma) \left(\beta \tilde{D} + (1-\beta) \tilde{Z} \right)\right] \geq 0,
\end{eqnarray*}
where $\beta=\frac{\lambda}{\lambda + \gamma}$. We can normalize $\lambda + \gamma$ to lie within the interval $[0,1]$.\footnote{If $\lambda + \gamma >1$, we can multiply each side of the inequality by $\frac{1}{\lambda + \gamma+1}$, and have $\frac{\lambda + \gamma}{\lambda + \gamma+1} \in [0,1]$.} By setting $\alpha=\lambda + \gamma$, this last inequality becomes
\begin{eqnarray*}
    \text{E}\left[Y_d \alpha \left(\beta \tilde{D} + (1-\beta) \tilde{Z} \right) \right] \geq 0.
\end{eqnarray*}

Hence, Inequality (\ref{eq.SDC1}) implies that, for any $(\alpha,\beta) \in [0,1]^2$, we have
\begin{eqnarray*}
    \text{E}\left[Y_d \alpha\left(\beta \tilde{D} + (1-\beta) \tilde{Z} \right)\right] \geq 0\ \text{  and  } \
    \text{E}\left[-Y_d \alpha\left(\beta \tilde{D} + (1-\beta) \tilde{Z} \right)\right] \leq 0.
\end{eqnarray*}
Intuitively, $\beta$ measures how tight is the constraint $\text{E}\left[Y_d\tilde{D}\right] \geq 0$ relatively to the constraint $ \text{E}\left[Y_d\tilde{Z}\right] \geq 0$, while $\alpha$ measures the extent to which the mixture of the constraints is binding. For example, when the constraint $\text{E}\left[Y_d\tilde{Z}\right] \geq 0$ is binding while the constraint $\text{E}\left[Y_d\tilde{D}\right] \geq 0$ is slack, then $\beta=0$, and the instrument $Z$ satisfies the zero covariance assumption, while the treatment variable $D$ is endogenous. In such a scenario, we expect the mixed constraint to be binding as well, that is $\alpha=1$. If instead, the treatment variable $D$ satisfies the zero covariance assumption, and the instrument $Z$ is endogenous, then  we expect $\beta=1$ and $\alpha=1$.

The latter inequalities are respectively equivalent to:\footnote{In the case where the $ATT/ATU$ is our parameter of interest, we would bound $\theta_{d|d'}\equiv \text{E}[Y_d\vert D=d']=\frac{\text{E}[Y_d\mathbbm{1}\{D=d'\}]}{\text{E}[\mathbbm{1}\{D=d'\}]}$. In such a case, we will equivalently write these inequalities as: \begin{eqnarray*}
    \text{E}\left[Y_d\left(\mathbbm{1}\{D=d'\}+ \alpha\left(\beta \tilde{D} + (1-\beta) \tilde{Z} \right)\right)\right] &\geq& \text{E}[Y_d \mathbbm{1}\{D=d'\}] \\
    \text{E}\left[Y_d\left(\mathbbm{1}\{D=d'\}- \alpha\left(\beta \tilde{D} + (1-\beta) \tilde{Z} \right)\right)\right] &\leq& \text{E}[Y_d \mathbbm{1}\{D=d'\}], 
\end{eqnarray*} 
and use the same technique we develop in this paper.}
\begin{eqnarray*}
    \text{E}\left[\delta^+_{S} Y_d \right] \geq \text{E}[Y_d ] \equiv \theta_d \ \text{  and  } \
    \text{E}\left[\delta^-_{S} Y_d\right] \leq \text{E}[Y_d] \equiv \theta_d,
\end{eqnarray*} 
where $\delta^+_{S} \equiv 1+ \alpha\left(\beta \tilde{D} + (1-\beta) \tilde{Z} \right)$ and $\delta^-_{S} \equiv 1- \alpha\left(\beta \tilde{D} + (1-\beta) \tilde{Z} \right)$.

Furthermore, using the identity $\mathbbm{1}\left\{D=d\right\} + \mathbbm{1}\left\{D\neq d\right\}=1$, we rewrite them as
\begin{eqnarray}
    \text{E}\left[\delta^+_{S} Y \mathbbm{1}\left\{D=d\right\} + \delta^+_{S} Y_d \mathbbm{1}\left\{D \neq d\right\} \right] &\geq& \theta_d,
    \label{eq.SDC11}
    \\
    \text{E}\left[\delta^-_{S} Y \mathbbm{1}\left\{D=d\right\} + \delta^-_{S} Y_d \mathbbm{1}\left\{D \neq d\right\} \right] &\leq&  \theta_d,
    \label{eq.SDC12}
\end{eqnarray}
respectively, given that $Y=Y_d$ when $D=d$.

Now, using Assumption BoS, we can bound the counterfactuals $\delta^+_{S} Y_d \mathbbm{1}\left\{D \neq d\right\}$ and $\delta^-_{S} Y_d \mathbbm{1}\left\{D \neq d\right\}$ as follows:
\begin{eqnarray*}
    \delta^+_{S} Y_d \mathbbm{1}\left\{D \neq d\right\}
    &\leq& \max\left\{ \delta^+_{S} \underline{y}_d, \delta^+_{S} \overline{y}_d \right\}\mathbbm{1}\left\{D \neq d\right\}
    \\
    \delta^-_{S} Y_d\mathbbm{1}\left\{D \neq d\right\}
    &\geq& \min\left\{ \delta^-_{S} \underline{y}_d,\delta^-_{S} \overline{y}_d \right\}\mathbbm{1}\left\{D \neq d\right\}.
\end{eqnarray*} 
Therefore, using inequalities (\ref{eq.SDC11}) and (\ref{eq.SDC12}), it follows that
\begin{eqnarray*}
    \text{E} \Big[ \overline{f}_d\left( Y, D, \delta^+_{S} \right) \Big] \geq \theta_d \text{  and  }
    \text{E} \Big[ \underline{f}_d\left( Y, D, \delta^-_{S} \right) \Big] \leq \theta_d
\end{eqnarray*}
for any $(\alpha, \beta) \in [0,1]^2$, where we define the function $\underline{f}_d$ and $\overline{f}_d$ as
\begin{eqnarray*}
    \underline{f}_d\left(Y, D, \delta \right) &\equiv&
    \mathbbm{1}\left\{D=d\right\}\delta Y + \mathbbm{1}\left\{D\neq d\right\} \min \left\{ \delta \underline{y}_d, \delta \overline{y}_d \right\}
    \\
    \overline{f}_d\left(Y, D, \delta \right) &\equiv&
    \mathbbm{1}\left\{D=d\right\}\delta Y + \mathbbm{1}\left\{D\neq d\right\} \max \left\{ \delta \underline{y}_d, \delta \overline{y}_d \right\}.
\end{eqnarray*}
We can then take the supremum and the infimum of the lower and upper bounds over $(\alpha, \beta)$, respectively, to obtain the following bounds for $\theta_d$:
\begin{equation*}
  I_{SDC1}^d \equiv \left[ \sup_{(\alpha, \beta) \in [0,1]^2} \text{E} \Big[ \underline{f}_d\left( Y, D, \delta^-_{S} \right) \Big], 
    \inf_{(\alpha, \beta) \in [0,1]^2} \text{E} \Big[ \overline{f}_d\left( Y, D, \delta^+_{S} \right) \Big] \right].
\end{equation*}
Similarly, using inequalities (\ref{eq.SDC2}), we derive the following bounds for $\theta_d$:
\begin{eqnarray*}
I_{SDC2}^d &\equiv& \left[  \sup_{(\alpha, \beta) \in [0,1]^2} \text{E} \Big[ \underline{f}_d\left( Y, D,\delta^+_{S} \right) \Big], \inf_{(\alpha, \beta) \in [0,1]^2} \text{E} \Big[ \overline{f}_d\left( Y, D, \delta^-_{S} \right) \Big] \right].
\end{eqnarray*}
All these results are summarized in the following proposition. 
\begin{proposition}\label{thm1}
Under Assumptions BoS and SDC, nonparametric bounds for the parameter $\theta_d$ are given by: 
\begin{equation*}
I_{SDC}^d \equiv I_{SDC1}^d  \cup  I_{SDC2}^d .
\end{equation*}
\end{proposition}
Proposition \ref{thm1} provides two-sided bounds on the potential outcome means, and then on the average treatment effects, which mainly relies on the bounded outcome assumption. \cite{Nevo2012} obtain two-sided bounds if $Cov(D,Z) <0$, and one-sided bounds if $Cov(D,Z) > 0$. We relax the parametric linear assumption at the expense of the bounded support assumption. In light of the following statement from the fourth paragraph of Section VI in \cite{Nevo2012} ``\textit{\ldots However, with a nonparametric functional, it is doubtful that our assumptions on the correlations of endogenous regressors and imperfect instruments with econometric errors would prove anywhere near as fruitful},'' we believe that the result of Proposition \ref{thm1} makes a positive contribution to the literature. However, we do not have a proof that the derived bounds are sharp at this point. We believe that this is an important theoretical question that can be investigated in future research. 

The bounds derived in Proposition \ref{thm1} are no wider than the usual Manski worst-case bounds without instrument, as the latter bounds are a special case of ours where $\alpha=0$. Furthermore, we expect the Manski bounds derived under the strict IV exogeneity condition, $\text{E}\left[Y_d\vert Z\right]=\text{E}\left[Y_d\right]$, to be weakly narrower than the bounds $I_{SDC}^d$. We provide a heuristic proof for this conjecture in the appendix. Intuitively, we expect the optimal value of $\beta$ to be equal to 0 under the strict IV exogeneity assumption, since the treatment variable $D$ is endogenous and the intrument $Z$ satisfies the zero covariance assumption. Using this restriction, we show that the lower (upper) bounds of $I_{SDC1}^d$ and $I_{SDC2}^d$ are each less (greater) than the Manski lower (upper) bound.

It is possible that the bounds in Proposition \ref{thm1} be empty. Indeed, if the supremum and the infimum in the bounds' expressions are attained at different values of $(\alpha, \beta)$, the lower bounds of $I_{SDC1}^d$ and $I_{SDC2}^d$ could be bigger than their respective upper bounds. When that happens for both bounds $I_{SDC1}^d$ and $I_{SDC2}^d$, we say that the model (Assumptions BoS and SDC) is rejected in the data. When only $I_{SDC1}^d$ ($I_{SDC2}^d$) is empty, then the assumption that the instrument $Z$ and the treatment $D$ are positively (negatively) correlated with the potential outcome $Y_d$ is rejected.

\subsection{Adding the less endogenous instrument assumption}
In this subsection, we combine Assumptions SDC and LEI in order to get tighter bounds on the parameter $\theta_d$.
Assumption LEI is equivalent to $\vert \frac{\text{E}\left[Y_d\tilde{D}\right]}{\sigma_D}\vert \geq \vert\frac{\text{E}\left[Y_d\tilde{Z}\right]}{\sigma_Z}\vert$.
Hence, Assumptions LEI and SDC imply that either of the followings is always true:
\begin{equation*}
    \frac{\text{E}\left[Y_d\tilde{D}\right]}{\sigma_D}
    \geq \frac{\text{E}\left[Y_d\tilde{Z}\right]}{\sigma_Z},
     \qquad \text{E}\left[Y_d\tilde{D}\right] \geq 0
    \qquad \textrm{and} \qquad \text{E}\left[Y_d\tilde{Z}\right] \geq 0,   
\end{equation*} 
or
\begin{equation*}
  \frac{\text{E}\left[Y_d\tilde{D}\right]}{\sigma_D}
    \leq \frac{\text{E}\left[Y_d\tilde{Z}\right]}{\sigma_Z},
     \qquad \text{E}\left[Y_d\tilde{D}\right] \leq 0
    \qquad \textrm{and} \qquad \text{E}\left[Y_d\tilde{Z}\right] \leq 0.   
\end{equation*} 
Differently, we can rewrite these inequalities as either
\begin{equation}
    \text{E}\left[Y_d \left(\tilde{D}\sigma_Z - \tilde{Z}\sigma_D\right)\right] \geq 0,
    \qquad \text{E}\left[Y_d\tilde{D}\right] \geq 0
    \qquad \textrm{and} \qquad \text{E}\left[Y_d\tilde{Z}\right] \geq 0,
    \label{eq.LEI1}
\end{equation} 
or
\begin{equation}
    \text{E}\left[Y_d \left(\tilde{D}\sigma_Z - \tilde{Z}\sigma_D\right)\right] \leq 0,
    \qquad \text{E}\left[Y_d\tilde{D}\right] \leq 0
    \qquad \textrm{and} \qquad \text{E}\left[Y_d\tilde{Z}\right] \leq 0.
    \label{eq.LEI2}
\end{equation}
Applying a similar reasoning as in the previous subsection to (\ref{eq.LEI1}), for any $(\alpha, \beta, \gamma) \in [0, 1]^3$ such that $1-\beta-\gamma \geq 0$, we have
\begin{eqnarray*}
    &\text{E}&\left[Y_d \alpha\left(\gamma \big(\tilde{D}\sigma_Z - \tilde{Z}\sigma_D\big) + \beta \tilde{D} + (1-\beta-\gamma) \tilde{Z} \right)\right] \geq 0\\
    \text{  and  } \
    &\text{E}&\left[-Y_d \alpha\left(\gamma \big(\tilde{D}\sigma_Z - \tilde{Z}\sigma_D\big) + \beta \tilde{D} + (1-\beta-\gamma) \tilde{Z} \right)\right] \leq 0.
\end{eqnarray*}
Since $\beta$ and $\gamma$ belong to a 2-simplex, we can parametrize $\gamma=(1-\beta)\mu$, where $\mu \in [0,1]$. Therefore, the above inequalities are equivalent to 
\begin{eqnarray*}
    \text{E}\left[\delta^+_{L} Y_d \right] \geq \text{E}[Y_d ] \equiv \theta_d \ \text{  and  } \
    \text{E}\left[\delta^-_{L} Y_d\right] \leq \text{E}[Y_d] \equiv \theta_d,
\end{eqnarray*} 
where $\delta^+_{L} \equiv 1+ \alpha\big((1-\beta)\mu (\tilde{D}\sigma_Z - \tilde{Z}\sigma_D) + \beta \tilde{D} + (1-\beta)(1-\mu) \tilde{Z} \big)$ and $\delta^-_{L} \equiv 1- \alpha\big((1-\beta)\mu  (\tilde{D}\sigma_Z - \tilde{Z}\sigma_D) + \beta \tilde{D} + (1-\beta)(1-\mu) \tilde{Z} \big)$.
Hence, we obtain the following bounds for $\theta_d$ from (\ref{eq.LEI1}):
\begin{equation*}
  I_{LEI1}^d \equiv \left[ \sup_{(\alpha, \beta, \mu) \in [0,1]^3} \text{E} \Big[ \underline{f}_d\left( Y, D, \delta^-_{L} \right) \Big], 
    \inf_{(\alpha, \beta, \mu) \in [0,1]^3} \text{E} \Big[ \overline{f}_d\left( Y, D, \delta^+_{L} \right) \Big] \right].
\end{equation*}
Likewise, we derive bounds for $\theta_d$ from (\ref{eq.LEI2}) as
\begin{equation*}
  I_{LEI2}^d \equiv \left[ \sup_{(\alpha, \beta, \mu) \in [0,1]^3} \text{E} \Big[ \underline{f}_d\left( Y, D, \delta^+_{L} \right) \Big], 
    \inf_{(\alpha, \beta, \mu) \in [0,1]^3} \text{E} \Big[ \overline{f}_d\left( Y, D, \delta^-_{L} \right) \Big] \right],
\end{equation*}
and the following proposition holds.
\begin{proposition}\label{thm2}
Under Assumptions BoS, SDC and LEI, nonparametric bounds for the parameter $\theta_d$ are given by: 
\begin{equation*}
I_{LEI}^d \equiv I_{LEI1}^d \cup I_{LEI2}^d.
\end{equation*}
\end{proposition}

The following corollary shows that $I_{LEI}^d$ is weakly tighter than $I_{SDC}^d$.
\begin{corollary}\label{cor1}
Under Assumptions BoS, SDC and LEI, we have
$$I_{LEI}^d \subseteq I_{SDC}^d.$$
\end{corollary}
The proof of Corollary \ref{cor1} follows from the fact that the bounds $I_{SDC}^d$ are a special case of the bounds $I_{LEI}^d$ where $\mu=0$.

\subsection{Adding the monotone treatment response assumption} In this subsection, we derive bounds on the potential outcome mean $\theta_d$ under Assumptions BoS, SDC, LEI, and MTR .
Under Assumption MTR, we have
\begin{eqnarray*}
    Y_d \mathbbm{1}\left\{D>d\right\} &=& Y_d \sum_{j=d+1}^{T} \mathbbm{1}\left\{D=j\right\}
    \, = \, \sum_{j=d+1}^{T} Y_d  \mathbbm{1}\left\{D=j\right\}
    \\
    &\leq& \sum_{j=d+1}^{T} Y  \mathbbm{1}\left\{D=j\right\}\, = \, Y \sum_{j=d+1}^{T}  \mathbbm{1}\left\{D=j\right\} \, = \, Y  \mathbbm{1}\left\{D>d\right\},
\end{eqnarray*}
where the inequality holds as
    $Y_d  \mathbbm{1}\left\{D=j\right\} \leq  Y_j  \mathbbm{1}\left\{D=j\right\}=Y \mathbbm{1}\left\{D=j\right\}$ for all $j>d$ for each $d$.
This result shows that under the MTR assumption, the counterfactual random variable $Y_d \mathbbm{1}\left\{D>d\right\}$ is bounded from above by the observed variable $Y  \mathbbm{1}\left\{D>d\right\}$. As we can see, this assumption considerably shrinks the upper bound on $Y_d \mathbbm{1}\left\{D>d\right\}$, which would be $\overline{y}_d \mathbbm{1}\left\{D>d\right\}$ otherwise under Assumption BoS.    
Similarly, we also have 
\begin{equation*}
    Y_d \mathbbm{1}\left\{D<d\right\} \quad \geq \quad Y  \mathbbm{1}\left\{D<d\right\}
\end{equation*}
for each $d$.
Without the MTR assumption, the lower bound on $Y_d \mathbbm{1}\left\{D<d\right\}$ would be $\underline{y}_d \mathbbm{1}\left\{D<d\right\}$ under Assumption BoS.
Combining these results together, the following inequalities hold under Assumptions BoS and MTR:
\begin{eqnarray*}
\begin{array}{lcl}
    \underline{y}_d  \mathbbm{1}\left\{D>d\right\} \quad \leq \quad &Y_d \mathbbm{1}\left\{D>d\right\}& \quad \leq \quad Y \mathbbm{1}\left\{D>d\right\},
    \\
    Y  \mathbbm{1}\left\{D<d\right\} \quad \leq \quad &Y_d \mathbbm{1}\left\{D<d\right\}& \quad \leq \quad \overline{y}_d \mathbbm{1}\left\{D<d\right\}.
 \end{array}
\end{eqnarray*}
Thus, for any $\delta \in \mathbb{R}$, we have the following bounds
\begin{eqnarray} 
\begin{array}{lcl}
    \min\{\delta \underline{y}_d, \delta Y \} \mathbbm{1}\left\{D>d\right\}
    \quad \leq  &\delta Y_d  \mathbbm{1}\left\{D>d\right\}& \leq \quad
    \max\{\delta \underline{y}_d, \delta Y \} \mathbbm{1}\left\{D>d\right\},
    \\ 
    \min\{\delta\overline{y}_d, \delta Y \} \mathbbm{1}\left\{D<d\right\}
    \quad \leq &\delta Y_d \mathbbm{1}\left\{D<d\right\}& \leq \quad
    \max\{\delta\overline{y}_d, \delta Y \} \mathbbm{1}\left\{D<d\right\}.
    \end{array}
    \label{eq.mon}
\end{eqnarray}
Moreover, as we have 
\begin{eqnarray*}
    Y_d \delta \mathbbm{1}\left\{D>d\right\}+Y_d \delta \mathbbm{1}\left\{D<d\right\}=Y_d \delta \mathbbm{1}\left\{D\neq d\right\},
\end{eqnarray*}
inequalities (\ref{eq.mon}) imply
\begin{eqnarray*}
    \delta Y_d\mathbbm{1}\left\{D\neq d\right\} &\leq& \max\left\{\delta\underline{y}_d, \delta Y\right\} \mathbbm{1}\left\{D>d\right\} + \max\left\{\delta\overline{y}_d, \delta Y\right\} \mathbbm{1}\left\{D<d\right\}, 
    \\
    \delta Y_d\mathbbm{1}\left\{D\neq d\right\} &\leq& \min\left\{\delta\underline{y}_d, \delta Y\right\} \mathbbm{1}\left\{D>d\right\} + \min\left\{\delta\overline{y}_d, \delta Y\right\} \mathbbm{1}\left\{D<d\right\},
\end{eqnarray*}
for any $\delta \in \mathbb{R}$.

Now, recall that inequalities (\ref{eq.LEI1}) implied by Assumptions SDC and LEI yield
\begin{eqnarray*}
    \text{E}\left[\delta^+_{L} Y \mathbbm{1}\left\{D=d\right\} + \delta^+_{L} Y_d \mathbbm{1}\left\{D \neq d\right\} \right] &\geq& \theta_d,
    \\
    \text{E}\left[\delta^-_{L} Y \mathbbm{1}\left\{D=d\right\} + \delta^-_{L} Y_d \mathbbm{1}\left\{D \neq d\right\} \right] &\leq&  \theta_d,
\end{eqnarray*}
for any $(\alpha, \beta, \mu) \in [0,1]^3$, and thus we have 
\begin{align*}
    \text{E} \Big[ \underline{m}_d\left( Y, D, \delta^-_{L} \right) \Big] \leq \theta_d \leq \text{E} \Big[ \overline{m}_d\left( Y, D, \delta^+_{L} \right) \Big]
\end{align*}
for any $(\alpha, \beta,\mu) \in [0,1]^3$, where we define the functions $\underline{m}_d$ and $\overline{m}_d$ as
\begin{eqnarray*}
    \underline{m}_d\left(Y, D, \delta \right) &\equiv&
    \mathbbm{1}\left\{D=d\right\}\delta Y + \mathbbm{1}\left\{D>d\right\} \min\{\delta\underline{y}_d, \delta Y \}  + \mathbbm{1}\left\{D<d\right\} \min\{\delta\overline{y}_d, \delta Y \}, 
    \\
    \overline{m}_d\left(Y, D, \delta \right) &\equiv&
    \mathbbm{1}\left\{D=d\right\}\delta Y + \mathbbm{1}\left\{D>d\right\}\max\{\delta\underline{y}_d, \delta Y \}  + \mathbbm{1}\left\{D<d\right\}\max\{\delta\overline{y}_d, \delta Y \} .  
\end{eqnarray*}
Likewise, inequalities (\ref{eq.LEI2}) under the MTR assumption yield the following implications for $\theta_d$:
\begin{align*}
    \text{E} \Big[ \underline{m}_d\left( Y, D, \delta^+_{L} \right) \Big] \leq \theta_d \leq \text{E} \Big[ \overline{m}_d\left( Y, D, \delta^-_{L} \right) \Big]
\end{align*}
for any $(\alpha, \beta,\mu) \in [0,1]^3$.
Therefore, we conclude that Assumptions BoS, SDC, LEI, and MTR together imply 
\begin{equation*}
    \theta_d \in  I_{MTR1}^d \cup I_{MTR2}^d \equiv I_{MTR}^d,
\end{equation*}
where
\begin{align*}
    I_{MTR1}^d &\equiv \left[ \sup_{(\alpha, \beta, \mu) \in [0,1]^3} \text{E} \Big[ \underline{m}_d\left( Y, D, \delta^-_{L} \right) \Big], \inf_{(\alpha, \beta, \mu) \in [0,1]^3} \text{E} \Big[ \overline{m}_d\left( Y, D, \delta^+_{L} \right) \Big] \right],\\
    I_{MTR2}^d &\equiv \left[ \sup_{(\alpha, \beta, \mu) \in [0,1]^3} \text{E} \Big[ \underline{m}_d\left( Y, D, \delta^+_{L} \right) \Big], \inf_{(\alpha, \beta, \mu) \in [0,1]^3} \text{E} \Big[ \overline{m}_d\left( Y, D, \delta^-_{L} \right) \Big] \right].
\end{align*}

Note that if we are not willing to impose Assumption LEI, the bounds can be obtained from $I_{MTR1}^d \cup I_{MTR2}^d$ using $\delta^+_{S}$ and $\delta^-_{S}$ instead of $\delta^+_{L}$ and $\delta^-_{L}$.

An important implication of the MTR assumption is that it weakly signs the ATE, i.e., $ATE(d,d')\geq 0$ for all $d > d'$. More precisely, bounds for $ATE(d,d')$, where $d>d'$, can be characterized as follows:
\begin{eqnarray*}
I_{MTR}^{ATE(d,d')}\equiv \left\{\theta^d-\theta^{d'}:\ \theta^d-\theta^{d'} \geq 0,\ \theta^d \in I_{MTR}^d \text{ and }  \theta^{d'} \in I_{MTR}^{d'}\right\}.
\end{eqnarray*}

\section{Inference}\label{inference}
\setcounter{equation}{0}
We want to construct confidence bounds for the set $I_{SDC}^d = I_{SDC1}^d  \cup  I_{SDC2}^d $. 
This is a special case of an intersection-union test as described in \cite{Berger1982}. We are going to construct confidence regions for the sets $I_{SDC1}^d$ and $I_{SDC2}^d$ using the intersection bounds framework of \cite{CLR2013} or \cite{AS2013}, and then take the union of the two confidence regions. \cite{Berger1996} showed that the union of the confidence regions has at least the same coverage rate as each confidence region. Identified sets with a similar structure have been considered in \cite{Chesher2020} and \cite{Machado2019}. As we explain in Section \ref{ident}, the sets $I_{SDC1}^d$ and $I_{SDC2}^d$ could each be empty, and may not overlap. However, in our empirical illustration below, they are nonempty and overlap in all cases.

We now explain how to rewrite the intersection bounds $ I_{SDC1}^d$ in such a way that it can be easily implemented using the \cite{CKLRstata} or \cite{AKS2017} Stata packages. Suppose that we draw two independent random variables $U_1$ and $U_2$ from the uniform distribution over $[0,1]$, independently of the data $(Y,D,Z)$. Then, we have
\begin{eqnarray*}
\text{E} \Big[ \underline{f}_d\big( Y, D, \delta^-_{S} \big) \Big\vert  U_1=\alpha, U_2=\beta \Big] = \text{E} \Big[ \underline{f}_d\big( Y, D, \delta^-_{S} \big) \Big],
\end{eqnarray*}
since $U_1$ and $U_2$ are independent of $(Y,D,Z)$. Therefore,
\begin{eqnarray*}
    I_{SDC1}^d &=& \Bigg[ \sup_{(\alpha, \beta) \in [0,1]^2} \text{E} \Big[ \underline{f}_d\big( Y, D, \delta^-_{S} \big) \Big\vert U_1=\alpha, U_2 = \beta \Big],\\
    && \qquad \qquad \qquad \qquad \qquad \inf_{(\alpha, \beta) \in [0,1]^2} \text{E} \Big[ \overline{f}_d\big( Y, D,\delta^+_{S} \big) \Big\vert U_1=\alpha, U_2 = \beta \Big] \Bigg].
\end{eqnarray*}

Hence, these bounds take the form of conditional moment inequalities, which can be implemented using existing inferential methods like \cite{CKLRstata} or \cite{AKS2017}. Similar results apply to the bounds $I_{LEI}^d$ and $I_{MTR}^d$, where there are three conditioning variables $U_1$, $U_2$, and $U_3$ instead of two. A similar technique has been proposed in \cite{Kedagni2018b} where they construct confidence sets for the potential outcome means under the IV zero-covariance assumption.

Using the \cite{CKLRstata} Stata package, we obtain confidence sets that asymptotically cover either the true parameter $\theta_d$ or the bounds for $\theta_d$ with pre-specified probability, through the \texttt{clr3bound} command.\footnote{The \texttt{clr2bound} command could be used to obtain confidence sets for the bounds for $\theta_d$ only.}

\section{Empirical illustration}\label{empirical}
\setcounter{equation}{0}
In this application, we use a data set drawn from the NLSYM. This data includes 3,010 young men who were ages 24-34 in 1976. It is the same data used in \cite{Card1995}. In our analysis, the outcome variable is log hourly wage in cents $(lwage)$, and the treatment variable is education $(educ)$ grouped in 4 categories: less than high school $(educ< 12\ years)$, high school $(12 \leq educ < 16)$, college degree $(16 \leq educ < 18)$, and graduate $(educ \geq 18)$.\footnote{As discussed in \cite{Andresen2021}, this discretization may induce some identification issues, and the results may be sensitive to it. For this reason, our empirical results should be seen as illustrative.}

Our imperfect IV is parental education. Since the work of \cite{Willis79}, parental education has been used as an IV. However, an individual's ability can be dependent on her parents' ability, which is correlated with parental education. For this reason, parents' education will not be a valid instrument. This fact is documented in \cite{Kedagni20}, who provided evidence that even after controlling for a measure of ability, parental education is not a good instrument. This result is contrary to the \cite{Lemke03} idea that controlling for some measure of child ability could make parental education a valid~IV. Nonetheless, it is reasonable to assume that parental education has the same sign of correlation with the individual's potential wage as the correlation between the person's potential wage and her own education.\footnote{The MIV and possibly MTS assumptions also seem reasonable in this empirical example. However, our goal in this section is to show how our derived bounds can be implemented in a real-world application.} It is also likely that parental education be less endogenous than is the person's own education. Finally, as in \cite{Manski2000}, we use the monotone treatment response assumption to tighten the bounds on the average returns to education. 

In theory, the outcome variable $lwage$ is unbounded. For practical reasons, we follow \cite{Ginther2000} to trim the log wage. The outcome variable that we use is defined as $Y=\tau$-quantile of $lwage$ if $lwage$ is less than or equal to its $\tau$-quantile, $Y=(1-\tau)$-quantile of $lwage$ if $lwage$ is greater than or equal to its $(1-\tau)$-quantile, and $Y=lwage$ otherwise. In our empirical illustration, we set $\tau=0.05$. We construct 95\% two-sided confidence sets on potential average log wages and their bounds using the \texttt{clr3bound} command of \cite{CKLRstata} in the Stata software. We estimate the conditional expectations using the parametric method, which is the default option for this Stata command. See the appendix for more details on the implementation.

We present the results with mother's education as an IIV. The results for father's education are in the appendix. Table \ref{Table:cs1} displays the 95\% confidence sets for the potential wage means and average returns to schooling under the SDC assumption, while Table \ref{Table:cs2} shows the confidence sets under both the SDC and LEI assumptions. Each table shows the confidence set for bounds on $\theta_d$ in the first two columns, the confidence set for the parameter $\theta_d$ in the third and forth columns, and the point estimates of the bounds in the fifth and sixth columns. The SDC+LEI bounds in Table \ref{Table:cs2} are generally narrower than the SDC bounds in Table \ref{Table:cs1}. This suggests that Assumption LEI provides some extra identifying power to the SDC assumption. However, the bounds seem wide and less informative in both cases. For example, the confidence set for $ATE(2,1) \equiv \theta_2-\theta_1$ under SDC+LEI is $[-0.98,1.02]$, implying that the average return to college degree compared to high school education varies between $-98\%$ and $102\%$, which is not very informative for an individual making a college decision. The corresponding point estimates of the bounds seem tighter $[-0.53, 0.69]$.

\begin{table}[ht] 
\caption{95\% Confidence regions for bounds and parameters under SDC}
\begin{center}
\begin{tabular}{l|cccccc} \label{Table:cs1}
           &   \multicolumn{2}{c}{For Bounds} & \multicolumn{2}{c}{For Parameters} & \multicolumn{2}{c}{Point Estimates} \\
Parameters &    Cf. LB & Cf. UB  &  Cf. LB & Cf. UB & $\,\,\,$ LB $\,\,\,$ & UB\\ \hline \hline
$\theta_0$ ($<$ high) &   	 5.51 &      6.91 &      5.53 &      6.89 &      5.65 &      6.70 \\ 
$\theta_1$ (high) &   		 5.86 &      6.64 &      5.88 &      6.62 &      5.97 &      6.45 \\ 
$\theta_2$ (college) &   	 5.62 &      6.94 &      5.64 &      6.92 &      5.77 &      6.76 \\ 
$\theta_3$ (graduate) &   	 5.51 &      7.01 &      5.53 &      6.99 &      5.64 &      6.85 \\ 
$\theta_0-\theta_1$   & 	 -1.13 &      1.05 &     -1.09 &      1.01 &     -0.81 &      0.73 \\
$\theta_2-\theta_1$ &   	 -1.02 &      1.08 &     -0.98 &      1.04 &     -0.68 &      0.79 \\
$\theta_3-\theta_1$ &  		 -1.13 &      1.15 &     -1.09 &      1.11 &     -0.81 &      0.88 \\
\hline\hline
\end{tabular}
\end{center}
\footnotesize \renewcommand{\baselineskip}{11pt} 
\textbf{Note:} Cf. LB (UB): lower (upper) bound of the 95\% confidence region; LB (UB): point estimate for the lower (upper) bound.
\end{table}

\begin{table}[ht] 
\caption{95\% Confidence regions for bounds and parameters under SDC and LEI}
\begin{center}
\begin{tabular}{l|cccccc} \label{Table:cs2}
           &   \multicolumn{2}{c}{For Bounds} & \multicolumn{2}{c}{For Parameters} & \multicolumn{2}{c}{Point Estimates} \\
Parameters &    Cf. LB & Cf. UB  &  Cf. LB & Cf. UB & $\,\,\,$ LB $\,\,\,$ & UB\\ \hline \hline
$\theta_0$ ($<$ high) &   	 5.52 &      6.91 &      5.54 &      6.89 &      5.70 &      6.64 \\ 
$\theta_1$ (high) &   		 5.86 &      6.64 &      5.88 &      6.62 &      6.01 &      6.36 \\ 
$\theta_2$ (college) &   	 5.62 &      6.92 &      5.64 &      6.90 &      5.83 &      6.70 \\ 
$\theta_3$ (graduate) &   	 5.51 &      7.00 &      5.53 &      6.98 &      5.69 &      6.78 \\ 
$\theta_0-\theta_1$   & 	 -1.12 &      1.05 &     -1.08 &      1.01 &     -0.66 &      0.63 \\
$\theta_2-\theta_1$ &   	 -1.02 &      1.06 &     -0.98 &      1.02 &     -0.53 &      0.69 \\
$\theta_3-\theta_1$ &  		 -1.13 &      1.14 &     -1.09 &      1.10 &     -0.67 &      0.77 \\
\hline\hline
\end{tabular}
\end{center}
\footnotesize \renewcommand{\baselineskip}{11pt} 
\textbf{Note:} Cf. LB (UB): lower (upper) bound of the 95\% confidence region; LB (UB): point estimate for the lower (upper) bound.
\end{table}
Furthermore, the confidence regions for the bounds and the parameters considerably shrink and become more informative when we add the MTR assumption (see Table \ref{Table:cs3}). We assume that the lower bound of $\theta_d$ is equal to the upper bound of $\theta_{d-1}$ whenever the former is less than the latter, because $Y_{d-1}$ cannot exceed $Y_d$ for each $d=1, 2, 3$ under the MTR assumption. Individuals with less than high school education could earn up to 94\% less than high school graduates ($ATE(0,1)$). Moreover, the confidence regions for $ATE(2, 1)$ and $ATE(3,1)$ suggest that college graduates could earn up to 55\% more than high school graduates, while individuals with a graduate degree earn between 41\% and 65\% higher wages than high school graduates (which approximately represents an annual return between 6.8\% and 10.8\%). Table II in \cite{Card2001} shows that the point estimates of the annual return to schooling in the US vary roughly between 5\% and 13\%. As we can see, our set estimates for the annual return are consistent with the existing range in the literature. 

If we impose the linear structure of \cite{Nevo2012} in this empirical exercise, we obtain the following point estimate bounds for $\theta$ under SDC and MTR: $[0, 0.16] \cup [0.24, +\infty)$. Indeed, the estimated correlation between $D$ and $Z$ is $0.3824>0$, and under SDC, the point estimate bounds for $\theta$ are $(-\infty, 0.16] \cup [0.24, +\infty)$. If one is willing to further assume that $Cov(D,U) \geq 0$, then the bounds for $\theta$ reduce to $[0, 0.16]$, which imply an annual return between 0 and 16\%. This range is consistent with the literature, but remains wide.

\begin{table}[ht] 
\caption{95\% Confidence regions for bounds and parameters under SDC, LEI, and MTR}
\begin{center}
\begin{tabular}{l|cccccc} \label{Table:cs3}
           &   \multicolumn{2}{c}{For Bounds} & \multicolumn{2}{c}{For Parameters} & \multicolumn{2}{c}{Point Estimates} \\
Parameters &    Cf. LB & Cf. UB  &  Cf. LB & Cf. UB & $\,\,\,$ LB $\,\,\,$ & UB\\ \hline \hline
$\theta_0$ ($<$ high) &   	 5.52 &      6.35 &      5.54 &      6.33 &      5.70 &      6.08 \\ 
$\theta_1$ (high) &   		 6.35 &      6.50 &      6.33 &      6.48 &      6.08 &      6.26 \\ 
$\theta_2$ (college) &   	 6.50 &      6.91 &      6.48 &      6.89 &      6.32 &      6.66 \\ 
$\theta_3$ (graduate) &   	 6.91 &      7.00 &      6.89 &      6.98 &      6.66 &      6.78 \\ 
$\theta_0-\theta_1$   & 	 -0.98 &      0.00 &     -0.94 &      0.00 &     -0.56 &      0.00 \\
$\theta_2-\theta_1$ &   	  0.00 &      0.55 &      0.00 &      0.55 &      0.06 &      0.58 \\
$\theta_3-\theta_1$ &  		  0.41 &      0.65 &      0.41 &      0.65 &      0.40 &      0.70 \\
\hline\hline
\end{tabular}
\end{center}
\footnotesize \renewcommand{\baselineskip}{11pt} 
\textbf{Note:} Cf. LB (UB): lower (upper) bound of the 95\% confidence region; LB (UB): point estimate for the lower (upper) bound.
\end{table}

\section{Conclusion}\label{conclusion}
In this paper, we derive nonparametric bounds on the average treatment effect when an imperfect instrument is available. We extend \posscite{Nevo2012} identification results to nonparametric models.  
We first assume that the sign of correlation between the imperfect instrument and the unobserved latent variables is the same as the correlation between the endogenous variable and the latent variables. 
We show that the MTS-MIV restrictions introduced by \cite{Manski2000, Manski2009}, jointly imply this assumption.
Second, we show how the assumption that the imperfect instrument is less endogenous than the treatment variable can help tighten the bounds. 
We also use the monotone treatment response assumption to get tighter bounds. The identified set takes the form of intersection bounds, which can be implemented using \posscite{CLR2013} inferential method. Finally, we illustrate our methodology using the National Longitudinal Survey of Young Men data to estimate returns to schooling.

\section*{Acknowledgements}
The authors are grateful to the Editor Petra Todd, and three anonymous referees for valuable suggestions and comments. They also thank Santiago Acerenza, Otavio Bartalotti, Helle Bunzel, Ismael Mourifi\'e, Vitor Possebom, and participants at the Iowa State econometrics workshop for helpful comments. All errors are ours.

\bibliographystyle{chicago}

\section{Appendix}
\subsection{Proofs of Results}

\textbf{Proof of Lemma 2.1:}
We have 
\begin{eqnarray*}
g_d^+(j)-g_d^-(j)\ &=& \sum^T_{\ell=j}\text{E}[Y_d\vert D=\ell]\frac{\mathbb P(D=\ell)}{\mathbb P(D\geq j)}-\sum^{j-1}_{\ell=1}\text{E}[Y_d\vert D=\ell]\frac{\mathbb P(D=\ell)}{\mathbb P(D < j)}
\end{eqnarray*}
Suppose without loss of generality that $\text{E}[Y_d \vert D=\ell]$ is increasing in $\ell$. Then 
\begin{eqnarray*}
\sum^T_{\ell=j}\text{E}[Y_d\vert D=\ell]\frac{\mathbb P(D=\ell)}{\mathbb P(D\geq j)} \geq \text{E}[Y_d\vert D=j],
\end{eqnarray*}
and 
\begin{eqnarray*}
 \sum^{j-1}_{\ell=1}\text{E}[Y_d\vert D=\ell]\frac{\mathbb P(D=\ell)}{\mathbb P(D < j)} \leq \text{E}[Y_d\vert D=j-1].
\end{eqnarray*}
Therefore
\begin{eqnarray*}
g_d^+(j)-g_d^-(j)\ &\geq& \text{E}[Y_d\vert D=j] -\text{E}[Y_d\vert D=j-1] \geq 0.
\end{eqnarray*}

On the other hand, we have
\begin{eqnarray*}
h^+_d(z) &=&  \text{E}\left[Y_d\vert Z\geq z\right] = \int^{\infty}_z \text{E}\left[Y_d\vert Z=v\right]\frac{f_Z(v)}{\mathbb P(Z\geq z)}dv,\\
&\geq& \int^{\infty}_z \text{E}\left[Y_d\vert Z=z\right]\frac{f_Z(v)}{\mathbb P(Z\geq z)}dv =\text{E}\left[Y_d\vert Z=z\right],
\end{eqnarray*}
where $f_Z$ is the density (or probability mass) of $Z$, and the inequality holds because $\text{E}\left[Y_d\vert Z=v\right]$ is increasing in $v$. 
Similarly, we have
\begin{eqnarray*}
h^-_d(z) &=& \text{E}\left[Y_d\vert Z\leq z\right] = \int^{z}_{-\infty} \text{E}\left[Y_d\vert Z=v\right]\frac{f_Z(v)}{\mathbb P(Z < z)}dv,\\
&\leq& \int^{z}_{-\infty} \text{E}\left[Y_d\vert Z=z\right]\frac{f_Z(v)}{\mathbb P(Z < z)}dv =\text{E}\left[Y_d\vert Z=z\right].
\end{eqnarray*}
Therefore
\begin{eqnarray*}
h^+_d(z) - h^-_d(z) &\geq& \text{E}\left[Y_d\vert Z=z\right]-\text{E}\left[Y_d\vert Z=z\right]=0.
\end{eqnarray*}
\hfill$\square$
\\

\textbf{Proof of Lemma 2.2:}
We first notice that 
\begin{eqnarray*}
D=\sum^T_{j=1} j \mathbbm{1}\{D=j\}=\sum^T_{j=1} j\left(\mathbbm{1}\{D\geq j\}-\mathbbm{1}\{D > j\}\right)=\sum^T_{j=1} \mathbbm{1}\{D\geq j\}.
\end{eqnarray*}
Then 
\begin{eqnarray*}
Cov(Y_d,D)=\sum^T_{j=1} Cov(Y_d,\mathbbm{1}\{D\geq j\}).
\end{eqnarray*}
When $Z$ is discrete, we have 
\begin{eqnarray*}
Cov(Y_d,Z)=\sum_{z \in \mathcal Z} Cov(Y_d,\mathbbm{1}\{Z\geq z\}).
\end{eqnarray*}
We show that 
\begin{eqnarray*}
Cov(Y_d,\mathbbm{1}\{D\geq j\}) &=& \mathbb P(D\geq j) \mathbb P(D<j) \left(\text{E}[Y_d \vert D\geq j]-\text{E}[Y_d\vert D < j]\right)\\
Cov(Y_d,\mathbbm{1}\{Z\geq z\}) &=& \mathbb P(Z\geq z) \mathbb P(Z<z) \left(\text{E}[Y_d \vert Z\geq z]-\text{E}[Y_d\vert Z < z]\right).
\end{eqnarray*}
From these results, it is straightforward to verify that if $Z$ is a binarized MTS-MIV for~$D$, then 
\begin{eqnarray*} 
Cov(Y_d,D)Cov(Y_d,Z)&=&\sum^T_{j=1} Cov(Y_d,\mathbbm{1}\{D\geq j\})\sum_{z \in \mathcal Z} Cov(Y_d,\mathbbm{1}\{Z\geq z\}),\\
&=& \sum^T_{j=1} \sum_{z \in \mathcal Z} Cov(Y_d,\mathbbm{1}\{D\geq j\}) Cov(Y_d,\mathbbm{1}\{Z\geq z\}),\\
&=& \sum^T_{j=1} \sum_{z \in \mathcal Z}  P(j,z)\big(g^+_d(j)-g^-_d(j)\big)\big(h^+_d(z)-h^-_d(z)\big) \geq 0,
\end{eqnarray*}
where $P(j,z)=\mathbb P(D\geq j) \mathbb P(D<j) \mathbb P(Z\geq z) \mathbb P(Z<z)$.

Now, consider the case where $Z$ is continuous. We are going to show that
\begin{eqnarray}
Cov(Y_d,Z)=\int^{\infty}_{-\infty} Cov(Y_d,\mathbbm{1}\{Z\geq z\})dz. \label{eq:cov1}
\end{eqnarray}

From the above equalities, we can show that if $Z$ is a binarized MTS-MIV for $D$, then 
\begin{eqnarray*} 
Cov(Y_d,D)Cov(Y_d,Z)&=&\sum^T_{j=1} Cov(Y_d,\mathbbm{1}\{D\geq j\})\int^{\infty}_{-\infty} Cov(Y_d,\mathbbm{1}\{Z\geq z\})dz,\\
&=& \sum^T_{j=1} \int^{\infty}_{-\infty} Cov(Y_d,\mathbbm{1}\{D\geq j\}) Cov(Y_d,\mathbbm{1}\{Z\geq z\})dz,\\
&=& \sum^T_{j=1} \int^{\infty}_{-\infty}  P(j,z)\big(g^+_d(j)-g^-_d(j)\big)\big(h^+_d(z)-h^-_d(z)\big)dz \geq 0.
\end{eqnarray*}

Now, let us prove (\ref{eq:cov1}). Let $(Y^{(1)}_d,Z^{(1)})$ and $(Y^{(2)}_d,Z^{(2)})$ be two independent copies of $(Y_d,Z)$. We have
\begin{eqnarray*}
Cov(Y_d,\mathbbm{1}\{Z\geq z\}) = \frac{1}{2}\text{E}[(Y^{(1)}_d-Y^{(2)}_d)(\mathbbm{1}\{Z^{(1)}\geq z\}-\mathbbm{1}\{Z^{(2)}\geq z\})].
\end{eqnarray*}
For any nonnegative random variable $X$, we know that 
\begin{eqnarray*}
X=\int^X_0  dx= \int^{\infty}_0 \mathbbm{1}\left\{X\geq x\right\}dx \text{  (layer cake representation)}. 
\end{eqnarray*}
Then 
\begin{eqnarray*}
Y_d&=&Y_d\mathbbm{1}\{Y_d\geq 0\}+ Y_d\mathbbm{1}\{Y_d < 0\}=Y_d\mathbbm{1}\{Y_d\geq 0\}- (-Y_d\mathbbm{1}\{Y_d < 0\}),\\
&=& \int^{\infty}_0 \mathbbm{1}\left\{Y_d\mathbbm{1}\{Y_d\geq 0\} \geq y\right\}dy - \int^{\infty}_0 \mathbbm{1}\left\{-Y_d\mathbbm{1}\{Y_d < 0\} \geq y\right\}dy.
\end{eqnarray*}
Notice that $\int^{\infty}_0 \mathbbm{1}\left\{Y_d\mathbbm{1}\{Y_d\geq 0\} \geq y\right\}dy=\int^{\infty}_0 \mathbbm{1}\left\{Y_d\geq y\right\}dy$ and $\int^{\infty}_0 \mathbbm{1}\left\{-Y_d\mathbbm{1}\{Y_d < 0\} \geq y\right\}dy=\int^{\infty}_0 \mathbbm{1}\left\{Y_d\leq -y\right\}dy$.
Therefore, 
\begin{eqnarray*}
Y_d&=&\int^{\infty}_0 \mathbbm{1}\left\{Y_d\geq y\right\}dy - \int^{\infty}_0 \mathbbm{1}\left\{Y_d\leq -y\right\}dy,\\
&=& \int^{\infty}_0 \mathbbm{1}\left\{Y_d\geq y\right\}dy + \int^{-\infty}_0 \mathbbm{1}\left\{Y_d\leq y'\right\}dy',\\
&=& \int^{\infty}_0 \mathbbm{1}\left\{Y_d\geq y\right\}dy - \int^0_{-\infty} \mathbbm{1}\left\{Y_d\leq y\right\}dy,  
\end{eqnarray*}
where the second equality holds from the change of variables $y'=-y$, and the last holds because $\int^0_{-\infty} \mathbbm{1}\left\{Y_d\leq y'\right\}dy'=\int^0_{-\infty} \mathbbm{1}\left\{Y_d\leq y\right\}dy$.
Thus,
\begin{eqnarray*}
Y_d^{(1)}-Y_d^{(2)} &=& \int^{\infty}_0 \mathbbm{1}\left\{Y_d^{(1)}\geq y\right\}-\mathbbm{1}\left\{Y_d^{(2)}\geq y\right\} dy\\
&& \qquad + \int^0_{-\infty} \mathbbm{1}\left\{Y_d^{(2)}\leq y\right\}-\mathbbm{1}\left\{Y_d^{(1)}\leq y\right\}dy,\\
&=&\int^{\infty}_0 \mathbbm{1}\left\{Y_d^{(2)}< y\right\}-\mathbbm{1}\left\{Y_d^{(1)} < y\right\} dy\\
&& \qquad + \int^0_{-\infty} \mathbbm{1}\left\{Y_d^{(2)}\leq y\right\}-\mathbbm{1}\left\{Y_d^{(1)}\leq y\right\}dy,,
\end{eqnarray*}
where the second equality holds from the identity $\mathbbm{1}\left\{X \geq x\right\}=1-\mathbbm{1}\left\{X < x\right\}$. For simplicity, suppose that $Y_d$ is continuous. Using the Fubini-Tonelli theorem, we can therefore rewrite $Cov(Y_d,\mathbbm{1}\{Z\geq z\})$ as follows:
\begin{eqnarray*}
Cov(Y_d,&\mathbbm{1}\{Z&\geq z\}) \\
 &=& \frac{1}{2} \int^{\infty}_{-\infty} \text{E}[(\mathbbm{1}\left\{Y_d^{(2)}\leq y\right\}-\mathbbm{1}\left\{Y_d^{(1)}\leq y\right\}) (\mathbbm{1}\{Z^{(1)}\geq z\}-\mathbbm{1}\{Z^{(2)}\geq z\})] dy,\\
&=& \frac{1}{2} \int^{\infty}_{-\infty} \text{E}[(\mathbbm{1}\left\{Y_d^{(2)}\leq y\right\}-\mathbbm{1}\left\{Y_d^{(1)}\leq y\right\}) (\mathbbm{1}\{Z^{(2)}\leq z\}-\mathbbm{1}\{Z^{(1)}\leq z\})] dy,\\
&=& \int^{\infty}_{-\infty} \text{E}[\mathbbm{1}\left\{Y_d \leq y, Z \leq z\right\}]-\text{E}[\mathbbm{1}\left\{Y_d \leq y\right\}] \text{E}[\mathbbm{1}\{Z\leq z\}] dy,\\
&=& \int^{\infty}_{-\infty} F_{(Y_d, Z)}(y, z) - F_{Y_d}(y)F_Z(z) dy
\end{eqnarray*}
where the second equality holds from the identity $\mathbbm{1}\left\{Z \geq z\right\}=1-\mathbbm{1}\left\{Z < z\right\}$ and the continuity of $Z$, $F_{(Y_d, Z)}$ is the joint cumulative distribution function of $(Y_d, Z)$, and $F_{Y_d}$ and $F_Z$ are the marginal cumulative distributions of $Y_d$ and $Z$, respectively.

From the Hoeffding's covariance identity, we have
\begin{eqnarray*}
	Cov(Y_d, Z) &=& \int^{\infty}_{-\infty}  \int^{\infty}_{-\infty} F_{(Y_d, Z)}(y, z) - F_{Y_d}(y)F_Z(z) dy dz,\\
	&=& \int^{\infty}_{-\infty} Cov(Y_d,\mathbbm{1}\{Z\geq z\})dz.
\end{eqnarray*}

Note that $Z$ the results still hold when $Z$ is mixed discrete-continuous.
\hfill$\square$
\\

\textbf{Proof that joint positive regression dependence (PRD) between $Y_d$ and $Z$, and between $Y_d$ and $D$ implies both joint MTS-MIV and joint positive quadrant dependence (PQD):}
Suppose that $\text{P}(Y_d > y \vert Z=z)$ is nondecreasing in $z$ for all $y$. Then $\text{P}(Y_d \leq y \vert Z=z)$ is decreasing in $z$, and we have
\begin{eqnarray*}
\text{P}(Y_d \leq y \vert Z < z) &=& \int^z_{-\infty} \text{P}(Y_d \leq y \vert Z=t)\frac{f_{Z}(t)}{\text{P}(Z < z)} dt,\\
&\geq& \int^z_{-\infty} \text{P}(Y_d \leq y \vert Z=t')\frac{f_{Z}(t)}{\text{P}(Z < z)} dt\ \text{ for all }\ t'\geq z,\\
&=& \text{P}(Y_d \leq y \vert Z=t') \ \text{ for all }\ t'\geq z.
\end{eqnarray*}
Therefore, by integrating both sides of the last inequality with respect to $t'$ over $[z,\infty)$, we have
\begin{eqnarray*}
\int^{\infty}_z \text{P}(Y_d \leq y \vert Z < z) \frac{f_{Z}(t')}{\text{P}(Z \geq z)} dt' &\geq& \int^{\infty}_z \text{P}(Y_d \leq y \vert Z=t') \frac{f_{Z}(t')}{\text{P}(Z \geq z)} dt',
\end{eqnarray*}
which is the same as
\begin{eqnarray*}
\text{P}(Y_d \leq y \vert Z < z) &\geq& \text{P}(Y_d \leq y \vert Z \geq z).
\end{eqnarray*}
From this last inequality, it follows successively:
\begin{eqnarray*}
\text{P}(Z \geq z) \text{P}(Y_d \leq y \vert Z < z) &\geq& \text{P}(Z \geq z) \text{P}(Y_d \leq y \vert Z \geq z),\\
(1-\text{P}(Z < z)) \text{P}(Y_d \leq y \vert Z < z) &\geq& \text{P}(Z \geq z) \text{P}(Y_d \leq y \vert Z \geq z),\\
\text{P}(Y_d \leq y \vert Z < z) &\geq& \text{P}(Z < z) \text{P}(Y_d \leq y \vert Z < z)  \\
&& \qquad + \, \text{P}(Z \geq z) \text{P}(Y_d \leq y \vert Z \geq z),\\
&=& \text{P}(Y_d \leq y).
\end{eqnarray*}
We conclude that positive regression dependence implies PQD.

On the other hand, $\text{P}(Y_d \leq y \vert Z=z) \geq \text{P}(Y_d \leq y \vert Z=z')$ for all $z' >z$ and all $y$, implies
\begin{eqnarray*}
\text{E}[Y_d \vert Z = z] &\leq& \text{E}[Y_d \vert Z = z'],
\end{eqnarray*}
which shows that MIV holds. Similarly, 
$\text{P}(Y_d \leq y \vert D=d) \geq \text{P}(Y_d \leq y \vert D=d')$ for all $d' >d$ implies MTS holds.
\hfill$\square$
\\

\textbf{Proof that joint PQD between $Y_d$ and $Z$, and between $Y_d$ and $D$ implies SDC:}
Suppose that $Y_d$ and $Z$ are PQD, and $Y_d$ and $D$ are also PQD. From the Hoeffding's covariance identity, we have
\begin{eqnarray*}
	Cov(Y_d, Z) = \int_{\mathbb R} \int_{\mathbb R} \Big( F_{(Y_d, Z)}(y, z) - F_{Y_d}(y)F_Z(z) \Big) dy dz,
\end{eqnarray*}
where $F_{(Y_d, Z)}$ is the joint cumulative distribution function of $(Y_d, Z)$, and $F_{Y_d}$ and $F_Z$ are the marginal cumulative distributions of $Y_d$ and $Z$, respectively.
We know that PDQ between $Y_d$ and $Z$ implies 
\begin{eqnarray*}
	F_{(Y_d, Z)}(y, z) \geq F_{Y_d}(y)F_Z(z)\ \text{ for all }\ y, z.
\end{eqnarray*}
Therefore, $Cov(Y_d, Z) \geq 0$. Similarly, $Cov(Y_d, D) \geq 0$. Hence, SDC holds.

\hfill$\square$
\\

\newpage

\textbf{Heuristic proof that the Manski bounds under IV mean independence (MI) are weakly narrower than our bounds under SDC:}
Recall that our bounds under SDC are given in Proposition 3.1 in the main text as
\begin{equation*}
I_{SDC}^d \equiv I_{SDC1}^d  \cup  I_{SDC2}^d, 
\end{equation*}
where
\begin{eqnarray*}
  I_{SDC1}^d &\equiv& \left[ \sup_{(\alpha, \beta) \in [0,1]^2} \text{E} \Big[ \underline{f}_d\left( Y, D, \delta^-_{S} \right) \Big], 
    \inf_{(\alpha, \beta) \in [0,1]^2} \text{E} \Big[ \overline{f}_d\left( Y, D, \delta^+_{S} \right) \Big] \right],\\
I_{SDC2}^d &\equiv& \left[  \sup_{(\alpha, \beta) \in [0,1]^2} \text{E} \Big[ \underline{f}_d\left( Y, D,\delta^+_{S} \right) \Big], \inf_{(\alpha, \beta) \in [0,1]^2} \text{E} \Big[ \overline{f}_d\left( Y, D, \delta^-_{S} \right) \Big] \right].
\end{eqnarray*}
where $\delta^+_{S} \equiv 1+ \alpha\left(\beta \tilde{D} + (1-\beta) \tilde{Z} \right)$, $\delta^-_{S} \equiv 1- \alpha\left(\beta \tilde{D} + (1-\beta) \tilde{Z} \right)$, and  the function $\underline{f}_d$ and $\overline{f}_d$ are defined as
\begin{eqnarray*}
    \underline{f}_d\left(Y, D, \delta \right) &\equiv&
    \mathbbm{1}\left\{D=d\right\}\delta Y + \mathbbm{1}\left\{D\neq d\right\} \min \left\{ \delta \underline{y}_d, \delta \overline{y}_d \right\}
    \\
    \overline{f}_d\left(Y, D, \delta \right) &\equiv&
    \mathbbm{1}\left\{D=d\right\}\delta Y + \mathbbm{1}\left\{D\neq d\right\} \max \left\{ \delta \underline{y}_d, \delta \overline{y}_d \right\}.
\end{eqnarray*}

Under the IV strict exogeneity assumption (mean independence), we conjecture that the optimal value of $\beta$ be equal to 0. Then, the lower and upper bounds for $I_{SDC1}^d$ respectively become:
\begin{eqnarray*}
&&\theta^\ell_{d, SDC1} \equiv \\
&& \qquad \sup_{\alpha \in [0,1]} \text{E} \Big[ \mathbbm{1}\left\{D=d\right\} Y \left(1- \alpha \tilde{Z}\right) + \mathbbm{1}\left\{D\neq d\right\} \min \left\{ \left(1- \alpha \tilde{Z} \right) \underline{y}_d, \left(1- \alpha \tilde{Z} \right) \overline{y}_d \right\} \Big],\\
&&\theta^u_{d, SDC1} \equiv \\
&&  \qquad \inf_{\alpha \in [0,1]} \text{E} \Big[ \mathbbm{1}\left\{D=d\right\} Y \left(1+ \alpha \tilde{Z}\right) + \mathbbm{1}\left\{D\neq d\right\} \max \left\{ \left(1+ \alpha \tilde{Z} \right) \underline{y}_d, \left(1+ \alpha \tilde{Z} \right) \overline{y}_d \right\} \Big]. 
\end{eqnarray*}
The Manski lower and upper bounds for $\theta_d$ under mean independence are respectively equal to:
\begin{eqnarray*}
\theta^\ell_{d, M} &\equiv& \sup_{z} \text{E} \Big[ \mathbbm{1}\left\{D=d\right\} Y + \mathbbm{1}\left\{D\neq d\right\} \underline{y}_d \vert Z=z\Big],\\
\theta^u_{d, M} &\equiv& \inf_{z} \text{E} \Big[ \mathbbm{1}\left\{D=d\right\} Y + \mathbbm{1}\left\{D\neq d\right\} \overline{y}_d \vert Z=z\Big]. 
\end{eqnarray*}
We are going to show $\theta^\ell_{d, SDC1} \leq \theta^\ell_{d, M}$, and $\theta^u_{d, SDC1} \geq \theta^u_{d, M}$. We have
\begin{eqnarray*}
&& \theta^\ell_{d, SDC1} \\
&=& \sup_{\alpha \in [0,1]} \Bigg\{ \text{E} \Big[ \left(\mathbbm{1}\left\{D=d\right\} Y  +  \underline{y}_d \mathbbm{1}\left\{D\neq d\right\}\right) \left(1- \alpha \tilde{Z}\right) \mathbbm{1} \left\{ 1- \alpha \tilde{Z} \geq 0 \right\} \Big]\\
&&\quad +\ \text{E} \Big[ \left(\mathbbm{1}\left\{D=d\right\} Y  +  \overline{y}_d \mathbbm{1}\left\{D\neq d\right\}\right) \left(1- \alpha \tilde{Z}\right) \mathbbm{1} \left\{ 1- \alpha \tilde{Z} < 0 \right\} \Big] \Bigg\},\\
 &=& \sup_{\alpha \in [0,1]} \Bigg\{ \text{E} \Big[ \text{E} \left[\mathbbm{1}\left\{D=d\right\} Y  +  \underline{y}_d \mathbbm{1}\left\{D\neq d\right\} \vert Z \right] \left(1- \alpha \tilde{Z}\right) \mathbbm{1} \left\{ 1- \alpha \tilde{Z} \geq 0 \right\} \Big]\\
&& \quad +\ \text{E} \Big[ \text{E} \left[\mathbbm{1}\left\{D=d\right\} Y  +  \overline{y}_d \mathbbm{1}\left\{D\neq d\right\} \vert Z \right] \left(1- \alpha \tilde{Z}\right) \mathbbm{1} \left\{ 1- \alpha \tilde{Z} < 0 \right\} \Big] \Bigg\},\\
 &\leq& \sup_{\alpha \in [0,1]} \Bigg\{ \text{E} \Big[ \sup_{z} \text{E} \left[\mathbbm{1}\left\{D=d\right\} Y  +  \underline{y}_d \mathbbm{1}\left\{D\neq d\right\} \vert Z=z\right] \left(1- \alpha \tilde{Z}\right) \mathbbm{1} \left\{ 1- \alpha \tilde{Z} \geq 0 \right\} \Big]\\
&& \quad +\ \text{E} \Big[\inf_{z} \text{E} \left[\mathbbm{1}\left\{D=d\right\} Y  +  \overline{y}_d \mathbbm{1}\left\{D\neq d\right\} \vert Z=z \right] \left(1- \alpha \tilde{Z}\right) \mathbbm{1} \left\{ 1- \alpha \tilde{Z} < 0 \right\} \Big] \Bigg\},\\
 &\leq& \sup_{\alpha \in [0,1]} \Bigg\{ \text{E} \Big[ \sup_{z} \text{E} \left[\mathbbm{1}\left\{D=d\right\} Y  +  \underline{y}_d \mathbbm{1}\left\{D\neq d\right\} \vert Z=z\right] \left(1- \alpha \tilde{Z}\right) \mathbbm{1} \left\{ 1- \alpha \tilde{Z} \geq 0 \right\} \Big]\\
&& \quad +\ \text{E} \Big[\sup_{z} \text{E} \left[\mathbbm{1}\left\{D=d\right\} Y  +  \underline{y}_d \mathbbm{1}\left\{D\neq d\right\} \vert Z=z \right] \left(1- \alpha \tilde{Z}\right) \mathbbm{1} \left\{ 1- \alpha \tilde{Z} < 0 \right\} \Big] \Bigg\},\\
 &=& \sup_{\alpha \in [0,1]} \Bigg\{\sup_{z} \text{E} \left[\mathbbm{1}\left\{D=d\right\} Y  +  \underline{y}_d \mathbbm{1}\left\{D\neq d\right\} \vert Z=z\right] \text{E} \Big[\left(1- \alpha \tilde{Z}\right) \mathbbm{1} \left\{ 1- \alpha \tilde{Z} \geq 0 \right\} \Big]\\
&& \quad +\ \sup_{z} \text{E} \left[\mathbbm{1}\left\{D=d\right\} Y  +  \underline{y}_d \mathbbm{1}\left\{D\neq d\right\} \vert Z=z \right]  \text{E} \Big[\left(1- \alpha \tilde{Z}\right) \mathbbm{1} \left\{ 1- \alpha \tilde{Z} < 0 \right\} \Big] \Bigg\},\\
 &=& \sup_{\alpha \in [0,1]} \Bigg\{\sup_{z} \text{E} \left[\mathbbm{1}\left\{D=d\right\} Y  +  \underline{y}_d \mathbbm{1}\left\{D\neq d\right\} \vert Z=z\right] \text{E} \Big[1- \alpha \tilde{Z} \Big] \Bigg\},\\
 &=& \sup_{z} \text{E} \left[\mathbbm{1}\left\{D=d\right\} Y  +  \underline{y}_d \mathbbm{1}\left\{D\neq d\right\} \vert Z=z\right]=\theta^\ell_{d, M},
\end{eqnarray*}
where the first equality holds from the definition of $\theta^\ell_{d, SDC1}$ and the identity $\mathbbm{1} \left\{ 1- \alpha \tilde{Z} \geq 0 \right\} +  \mathbbm{1} \left\{ 1- \alpha \tilde{Z} < 0 \right\}=1$, the second equality holds from the law of iterated expectations, the first inequality holds from the following inequalities 
\begin{eqnarray*}
\text{E} \left[\mathbbm{1}\left\{D=d\right\} Y  +  \underline{y}_d \mathbbm{1}\left\{D\neq d\right\} \vert Z\right] &\leq& \sup_{z} \text{E} \left[\mathbbm{1}\left\{D=d\right\} Y  +  \underline{y}_d \mathbbm{1}\left\{D\neq d\right\} \vert Z=z\right],\\
  \text{E} \left[\mathbbm{1}\left\{D=d\right\} Y  +  \overline{y}_d \mathbbm{1}\left\{D\neq d\right\} \vert Z\right] &\geq& \inf_{z} \text{E} \left[\mathbbm{1}\left\{D=d\right\} Y  +  \overline{y}_d \mathbbm{1}\left\{D\neq d\right\} \vert Z=z\right],
\end{eqnarray*}
the second inequality follows from the fact Manski's lower bound is less than his upper bound under mean independence, the third equality holds from the fact that the quantity $\sup_{z} \text{E} \left[\mathbbm{1}\left\{D=d\right\} Y  +  \underline{y}_d \mathbbm{1}\left\{D\neq d\right\} \vert Z=z\right]$ is constant, the fourth equality holds from this identity $\mathbbm{1} \left\{ 1- \alpha \tilde{Z} \geq 0 \right\} +  \mathbbm{1} \left\{ 1- \alpha \tilde{Z} < 0 \right\}=1$, and the fifth equality holds from $\text{E} \Big[1- \alpha \tilde{Z} \Big]=1$ since $\text{E} \Big[\tilde{Z} \Big]=0$.

Similarly, we have
\begin{eqnarray*}
&&\theta^u_{d, SDC1}\\
 &=& \inf_{\alpha \in [0,1]} \Bigg\{ \text{E} \Big[ \left(\mathbbm{1}\left\{D=d\right\} Y  +  \overline{y}_d \mathbbm{1}\left\{D\neq d\right\}\right) \left(1+ \alpha \tilde{Z}\right) \mathbbm{1} \left\{ 1+ \alpha \tilde{Z} \geq 0 \right\} \Big]\\
&&\quad +\ \text{E} \Big[ \left(\mathbbm{1}\left\{D=d\right\} Y  +  \underline{y}_d \mathbbm{1}\left\{D\neq d\right\}\right) \left(1+ \alpha \tilde{Z}\right) \mathbbm{1} \left\{ 1+ \alpha \tilde{Z} < 0 \right\} \Big] \Bigg\},\\
 &=& \inf_{\alpha \in [0,1]} \Bigg\{ \text{E} \Big[ \text{E} \left[\mathbbm{1}\left\{D=d\right\} Y  +  \overline{y}_d \mathbbm{1}\left\{D\neq d\right\} \vert Z \right] \left(1+ \alpha \tilde{Z}\right) \mathbbm{1} \left\{ 1+ \alpha \tilde{Z} \geq 0 \right\} \Big]\\
&&\quad  +\ \text{E} \Big[ \text{E} \left[\mathbbm{1}\left\{D=d\right\} Y  +  \underline{y}_d \mathbbm{1}\left\{D\neq d\right\} \vert Z \right] \left(1+ \alpha \tilde{Z}\right) \mathbbm{1} \left\{ 1+ \alpha \tilde{Z} < 0 \right\} \Big] \Bigg\},\\
 &\geq& \inf_{\alpha \in [0,1]} \Bigg\{ \text{E} \Big[ \inf_{z} \text{E} \left[\mathbbm{1}\left\{D=d\right\} Y  +  \overline{y}_d \mathbbm{1}\left\{D\neq d\right\} \vert Z=z\right] \left(1+ \alpha \tilde{Z}\right) \mathbbm{1} \left\{ 1+ \alpha \tilde{Z} \geq 0 \right\} \Big]\\
&&\quad  +\ \text{E} \Big[\sup_{z} \text{E} \left[\mathbbm{1}\left\{D=d\right\} Y  +  \underline{y}_d \mathbbm{1}\left\{D\neq d\right\} \vert Z=z \right] \left(1+ \alpha \tilde{Z}\right) \mathbbm{1} \left\{ 1+ \alpha \tilde{Z} < 0 \right\} \Big] \Bigg\},\\
 &\geq& \inf_{\alpha \in [0,1]} \Bigg\{ \text{E} \Big[ \inf_{z} \text{E} \left[\mathbbm{1}\left\{D=d\right\} Y  +  \overline{y}_d \mathbbm{1}\left\{D\neq d\right\} \vert Z=z\right] \left(1+ \alpha \tilde{Z}\right) \mathbbm{1} \left\{ 1+ \alpha \tilde{Z} \geq 0 \right\} \Big]\\
&&\quad  +\ \text{E} \Big[\inf_{z} \text{E} \left[\mathbbm{1}\left\{D=d\right\} Y  +  \overline{y}_d \mathbbm{1}\left\{D\neq d\right\} \vert Z=z \right] \left(1+ \alpha \tilde{Z}\right) \mathbbm{1} \left\{ 1+ \alpha \tilde{Z} < 0 \right\} \Big] \Bigg\},\\
 &=& \inf_{\alpha \in [0,1]} \Bigg\{\inf_{z} \text{E} \left[\mathbbm{1}\left\{D=d\right\} Y  +  \overline{y}_d \mathbbm{1}\left\{D\neq d\right\} \vert Z=z\right] \text{E} \Big[\left(1+ \alpha \tilde{Z}\right) \mathbbm{1} \left\{ 1+ \alpha \tilde{Z} \geq 0 \right\} \Big]\\
&&\quad +\ \inf_{z} \text{E} \left[\mathbbm{1}\left\{D=d\right\} Y  +  \overline{y}_d \mathbbm{1}\left\{D\neq d\right\} \vert Z=z \right]  \text{E} \Big[\left(1+ \alpha \tilde{Z}\right) \mathbbm{1} \left\{ 1+ \alpha \tilde{Z} < 0 \right\} \Big] \Bigg\},\\
 &=& \inf_{\alpha \in [0,1]} \Bigg\{\inf_{z} \text{E} \left[\mathbbm{1}\left\{D=d\right\} Y  +  \overline{y}_d \mathbbm{1}\left\{D\neq d\right\} \vert Z=z\right] \text{E} \Big[1+ \alpha \tilde{Z} \Big] \Bigg\},\\
 &=& \inf_{z} \text{E} \left[\mathbbm{1}\left\{D=d\right\} Y  +  \overline{y}_d \mathbbm{1}\left\{D\neq d\right\} \vert Z=z\right]=\theta^u_{d, M},
\end{eqnarray*}
where the first equality holds from the definition of $\theta^u_{d, SDC1}$ and the identity $\mathbbm{1} \left\{ 1+ \alpha \tilde{Z} \geq 0 \right\} +  \mathbbm{1} \left\{ 1+ \alpha \tilde{Z} < 0 \right\}=1$, the second equality holds from the law of iterated expectations, the first inequality holds from the following inequalities 
\begin{eqnarray*}
\text{E} \left[\mathbbm{1}\left\{D=d\right\} Y  +  \underline{y}_d \mathbbm{1}\left\{D\neq d\right\} \vert Z\right] &\leq& \sup_{z} \text{E} \left[\mathbbm{1}\left\{D=d\right\} Y  +  \underline{y}_d \mathbbm{1}\left\{D\neq d\right\} \vert Z=z\right],\\
  \text{E} \left[\mathbbm{1}\left\{D=d\right\} Y  +  \overline{y}_d \mathbbm{1}\left\{D\neq d\right\} \vert Z\right] &\geq& \inf_{z} \text{E} \left[\mathbbm{1}\left\{D=d\right\} Y  +  \overline{y}_d \mathbbm{1}\left\{D\neq d\right\} \vert Z=z\right],
\end{eqnarray*}
the second inequality follows from the fact Manski's lower bound is less than his upper bound under mean independence, the third equality holds from the fact that the quantity $\inf_{z} \text{E} \left[\mathbbm{1}\left\{D=d\right\} Y  +  \overline{y}_d \mathbbm{1}\left\{D\neq d\right\} \vert Z=z\right]$ is constant, the fourth equality holds from this identity $\mathbbm{1} \left\{ 1+ \alpha \tilde{Z} \geq 0 \right\} +  \mathbbm{1} \left\{ 1+ \alpha \tilde{Z} < 0 \right\}=1$, and the fifth equality holds from $\text{E} \Big[1+ \alpha \tilde{Z} \Big]=1$ since $\text{E} \Big[\tilde{Z} \Big]=0$.

Analogously, we can show similar results for the lower and upper bounds of $I_{SDC2}^d$.

\hfill$\square$ \\

\subsection{Implementation of the bounds}\label{apx:impl}
In this section, we show the different steps for implementing the method developed in the paper. We want to construct confidence bounds on the parameter $\theta_d \in I_{SDC1}^d$, where
\begin{eqnarray*}
    I_{SDC1}^d &=& \Bigg[ \sup_{(\alpha, \beta) \in [0,1]^2} \text{E} \Big[ \underline{f}_d\big( Y, D, \delta^-_{S} \big) \Big\vert U_1=\alpha, U_2 = \beta \Big],\\
    && \qquad \qquad \qquad \qquad \qquad \inf_{(\alpha, \beta) \in [0,1]^2} \text{E} \Big[ \overline{f}_d\big( Y, D,\delta^+_{S} \big) \Big\vert U_1=\alpha, U_2 = \beta \Big] \Bigg].
\end{eqnarray*}
As explained in the main text, the implementation can be done using the Stata package developed by \cite{CKLRstata}. Assuming that we have an i.i.d. data with finite moments, the conditions for the implementation of the \cite{CLR2013} method are likely to hold in our framework. We provide below the code for the implementation.\\

\lstset{basicstyle=\small}
\begin{lstlisting}
set more off
gen Y = lwage
centile(Y), centile(5 95)
replace Y = r(c_1) if Y < r(c_1)
replace Y = r(c_2) if Y > r(c_2)
replace educ = 0 if educ < 12
replace educ = 1 if educ >= 12 & educ < 16
replace educ = 2 if educ >= 16 & educ < 18
replace educ = 3 if educ >= 18
replace motheduc = 0 if motheduc < 12
replace motheduc = 1 if motheduc >= 12 & motheduc < 16
replace motheduc = 2 if motheduc >= 16 & motheduc < 18
replace motheduc = 3 if motheduc >= 18 & motheduc != .
gen D = (educ==0)
gen Z = motheduc
sum Y if D == 1
scalar Y1up = r(max)
scalar Y1lo = r(min)
sum Y if D == 0
scalar Y0up = r(max)
scalar Y0lo = r(min)
sum Z
scalar EZ = r(mean)
sum educ
scalar ED = r(mean)
set seed 12345
gen Alpha=runiform(0,1)
gen Beta=runiform(0,1)
gen RAlpha = floor((_n-1)/11)/10
gen RBeta = mod(_n-1, 11)/10
replace RAlpha=. if RAlpha>1
replace RBeta=. if RAlpha>1
gen del_minus = 1 - Alpha*(Beta*(educ-ED) + (1-Beta)*(Z-EZ))
gen del_plus = 1 + Alpha*(Beta*(educ-ED) + (1-Beta)*(Z-EZ))
gen ldepen1 = Y*D*del_minus
    + min(Y1lo*(1-D)*del_minus,Y1up*(1-D)*del_minus)
gen ldepen2 = Y*D*del_plus
    + min(Y1lo*(1-D)*del_plus,Y1up*(1-D)*del_plus)
gen udepen1 = Y*D*del_plus
    + max(Y1lo*(1-D)*del_plus,Y1up*(1-D)*del_plus)
gen udepen2 = Y*D*del_minus
    + max(Y1lo*(1-D)*del_minus,Y1up*(1-D)*del_minus)
clr3bound ((ldepen1 (Alpha Beta) (RAlpha RBeta))) 
    ((udepen1 (Alpha Beta) (RAlpha RBeta)))
    , level(0.95) norseed rnd(20000)
clr3bound ((ldepen2 (Alpha Beta) (RAlpha RBeta))) 
    ((udepen2 (Alpha Beta) (RAlpha RBeta)))
    , level(0.95) norseed rnd(20000)
\end{lstlisting}

\subsection{Additional empirical results}\label{appx2}

\subsubsection{Additional results using mother's education as an IIV} Table \ref{Table:cs5} shows additional results using mother's education as an imperfect IV. Compared to Table \ref{Table:cs3}, the bounds are generally wider here without Assumption LEI, suggesting that the additional inequalities from Assumption LEI are binding.

\begin{table}[ht] 
\caption{95\% Confidence regions for bounds and parameters under SDC and MTR}
\begin{center}
\begin{tabular}{l|cccccc} \label{Table:cs5}
           &   \multicolumn{2}{c}{For Bounds} & \multicolumn{2}{c}{For Parameters} & \multicolumn{2}{c}{Point Estimates} \\
Parameters &    Cf. LB & Cf. UB  &  Cf. LB & Cf. UB & $\,\,\,$ LB $\,\,\,$ & UB\\ \hline \hline
$\theta_0$ ($<$ high) &   	 5.51 &      6.35 &      5.53 &      6.33 &      5.65 &      6.15 \\ 
$\theta_1$ (high) &   		 6.35 &      6.50 &      6.33 &      6.48 &      6.15 &      6.34 \\ 
$\theta_2$ (college) &   	 6.50 &      6.92 &      6.48 &      6.90 &      6.34 &      6.73 \\ 
$\theta_3$ (graduate) &   	 6.92 &      7.01 &      6.90 &      6.99 &      6.73 &      6.85 \\ 
$\theta_0-\theta_1$   & 	 -0.99 &      0.00 &     -0.95 &      0.00 &     -0.69 &      0.00 \\
$\theta_2-\theta_1$ &   	  0.00 &      0.56 &      0.00 &      0.56 &      0.00 &      0.58 \\
$\theta_3-\theta_1$ &  		  0.42 &      0.66 &      0.42 &      0.66 &      0.40 &      0.70 \\
\hline\hline
\end{tabular}
\end{center}
\footnotesize \renewcommand{\baselineskip}{11pt} 
\textbf{Note:} Cf. LB (UB): lower (upper) bound of the 95\% confidence region; LB (UB): point estimate for the lower (upper) bound.
\end{table}

\subsubsection{Results using father's education as an IIV} \label{apx:fatheduc}
Tables \ref{Table:cs6}, \ref{Table:cs7} and \ref{Table:cs8} present the results using father's education as an imperfect IV. The results in Table \ref{Table:cs6} are similar to those we obtain using mother's education as an IIV. Tables \ref{Table:cs7} and \ref{Table:cs8} display the results with race as a control variable; Table \ref{Table:cs7} shows the results with \textit{black} = 1 while Table \ref{Table:cs8} shows the results with \textit{black} = 0.
\begin{table}[ht] 
\caption{95\% Confidence regions for bounds and parameters under SDC, LEI and MTR using father's education as IIV}
\begin{center}
\begin{tabular}{l|cccccc} \label{Table:cs6}
           &   \multicolumn{2}{c}{For Bounds} & \multicolumn{2}{c}{For Parameters} & \multicolumn{2}{c}{Point Estimates} \\
Parameters &    Cf. LB & Cf. UB  &  Cf. LB & Cf. UB & $\,\,\,$ LB $\,\,\,$ & UB\\ \hline \hline
$\theta_0$ ($<$ high) &   	 5.50 &      6.40 &      5.52 &      6.38 &      5.63 &      6.20 \\ 
$\theta_1$ (high) &   		 6.40 &      6.52 &      6.38 &      6.50 &      6.20 &      6.33 \\ 
$\theta_2$ (college) &   	 6.52 &      6.92 &      6.50 &      6.90 &      6.33 &      6.77 \\ 
$\theta_3$ (graduate) &   	 6.92 &      7.02 &      6.90 &      7.00 &      6.77 &      6.83 \\ 
$\theta_0-\theta_1$   & 	 -1.02 &      0.00 &     -0.98 &      0.00 &     -0.70 &      0.00 \\
$\theta_2-\theta_1$ &   	  0.00 &      0.52 &      0.00 &      0.52 &      0.00 &      0.56 \\
$\theta_3-\theta_1$ &  		  0.40 &      0.63 &      0.40 &      0.63 &      0.43 &      0.63 \\
\hline\hline
\end{tabular}
\end{center}
\footnotesize \renewcommand{\baselineskip}{11pt} 
\textbf{Note:} Cf. LB (UB): lower (upper) bound of the 95\% confidence region; LB (UB): point estimate for the lower (upper) bound.
\end{table}

\begin{table}[ht] 
\caption{95\% Confidence regions for bounds and parameters under SDC, LEI and MTR using father's education as IIV with \textit{black} = 1}
\begin{center}
\begin{tabular}{l|cccccc} \label{Table:cs7}
           &   \multicolumn{2}{c}{For Bounds} & \multicolumn{2}{c}{For Parameters} & \multicolumn{2}{c}{Point Estimates} \\
Parameters &    Cf. LB & Cf. UB  &  Cf. LB & Cf. UB & $\,\,\,$ LB $\,\,\,$ & UB\\ \hline \hline
$\theta_0$ ($<$ high) &   	 5.36 &      6.34 &      5.38 &      6.32 &      \textbf{empty} &       \textbf{empty} \\ 
$\theta_1$ (high) &   		 6.34 &      6.62 &      6.32 &      6.60 &       \textbf{empty} &       \textbf{empty} \\ 
$\theta_2$ (college) &   	 6.62 &      7.12 &      6.60 &      7.10 &      6.28 &      6.62 \\ 
$\theta_3$ (graduate) &   	 7.12 &      7.20 &      7.10 &      7.18 &       \textbf{empty} &       \textbf{empty} \\ 
$\theta_0-\theta_1$   & 	 -1.26 &      0.00 &     -1.22 &      0.00 &      - &      - \\
$\theta_2-\theta_1$ &   	  0.00 &      0.78 &      0.00 &      0.78 &      - &       - \\
$\theta_3-\theta_1$ &  		  0.50 &      0.86 &      0.50 &      0.86 &      - &       - \\
\hline\hline
\end{tabular}
\end{center}
\footnotesize \renewcommand{\baselineskip}{11pt} 
\textbf{Note:} Cf. LB (UB): lower (upper) bound of the 95\% confidence region; LB (UB): point estimate for the lower (upper) bound.
\end{table}

\begin{table}[ht] 
\caption{95\% Confidence regions for bounds and parameters under SDC, LEI and MTR using father's education as IIV with \textit{black} = 0}
\begin{center}
\begin{tabular}{l|cccccc} \label{Table:cs8}
           &   \multicolumn{2}{c}{For Bounds} & \multicolumn{2}{c}{For Parameters} & \multicolumn{2}{c}{Point Estimates} \\
Parameters &    Cf. LB & Cf. UB  &  Cf. LB & Cf. UB & $\,\,\,$ LB $\,\,\,$& UB\\ \hline \hline
$\theta_0$ ($<$ high)    	&5.49 &      6.41 &      5.51 &      6.39 &       5.81 &      6.04 \\ 
$\theta_1$ (high)    		&6.41 &      6.53 &      6.39 &      6.51 &       \textbf{empty} &      \textbf{empty} \\ 
$\theta_2$ (college)    	&6.53 &      6.92 &      6.51 &      6.90 &      6.49 &      6.54 \\ 
$\theta_3$ (graduate)    	&6.92 &      7.04 &      6.90 &      7.02 &      6.61 &      6.63 \\ 
$\theta_0-\theta_1$    		&-1.04 &      0.00 &     -1.00 &      0.00 &      - &       - \\
$\theta_2-\theta_1$    		& 0.00 &      0.51 &      0.00 &      0.51 &      - &       - \\
$\theta_3-\theta_1$   		& 0.40 &      0.63 &      0.40 &      0.63 &      - &       - \\
\hline\hline
\end{tabular}
\end{center}
\footnotesize \renewcommand{\baselineskip}{11pt} 
\textbf{Note:} Cf. LB (UB): lower (upper) bound of the 95\% confidence region; LB (UB): point estimate for the lower (upper) bound.
\end{table}


\subsection{An example where binarized MTS-MIV holds, but MTS-MIV fails}\label{ex:app}

\begin{figure}[ht]
    \includegraphics[width=0.8\textwidth]{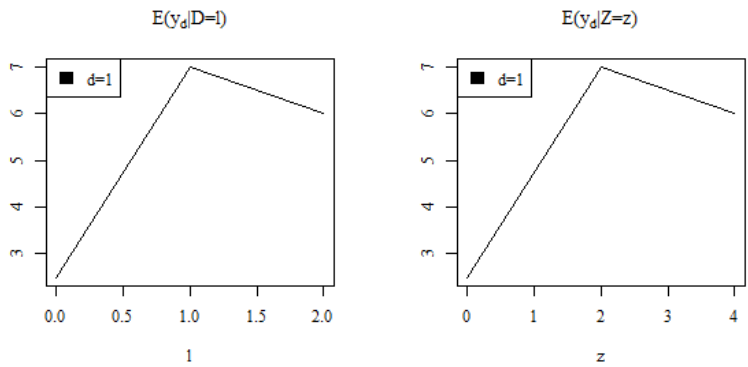}
    \centering
    \caption{Numerical illustration of a violation of MTS and MIV}
    \label{fig.ey1.exam0921}
\end{figure}

\begin{figure}[ht]
    \includegraphics[width=0.7\textwidth]{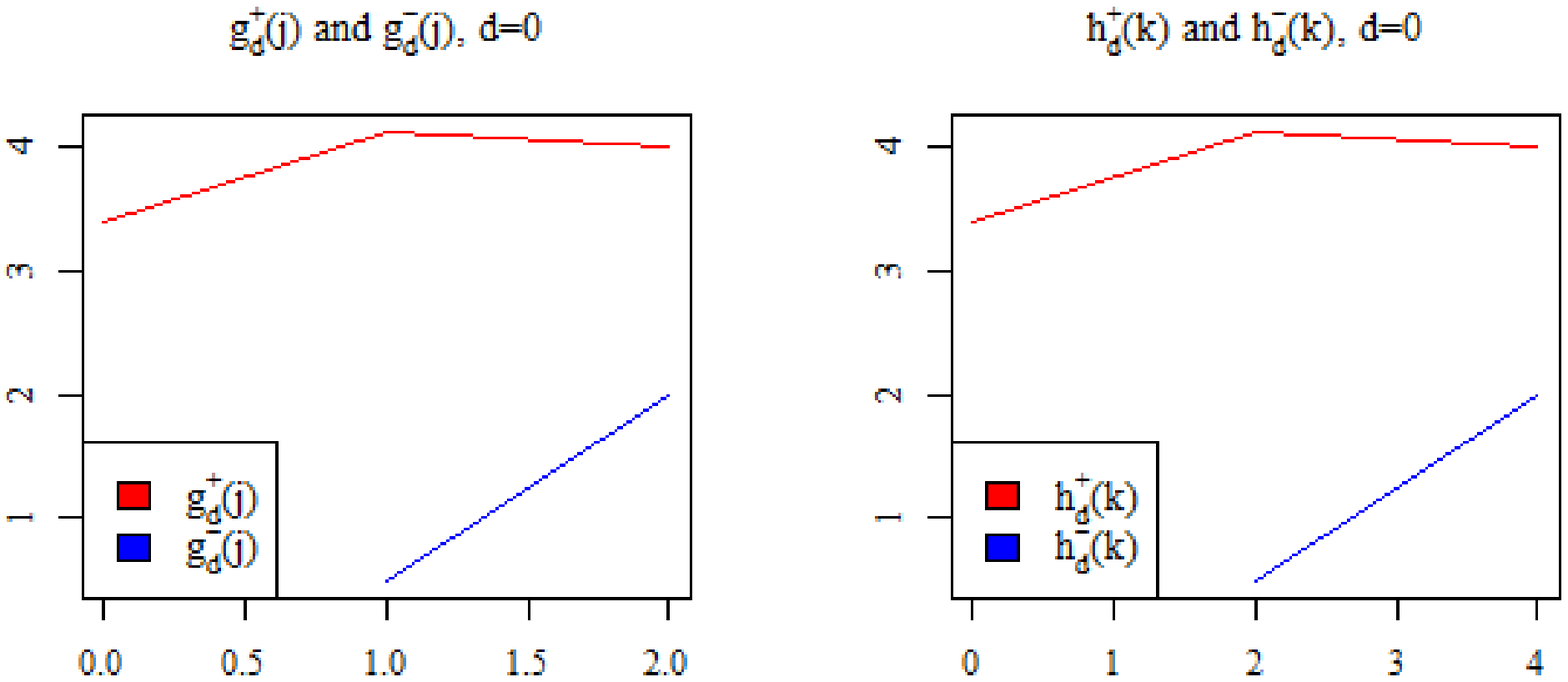}
    \centering
    \caption{Numerical illustration of binarized MTS-MIV 1}
    \label{fig.gh1.exam0921}
\end{figure}

\begin{figure}[ht]
    \includegraphics[width=0.7\textwidth]{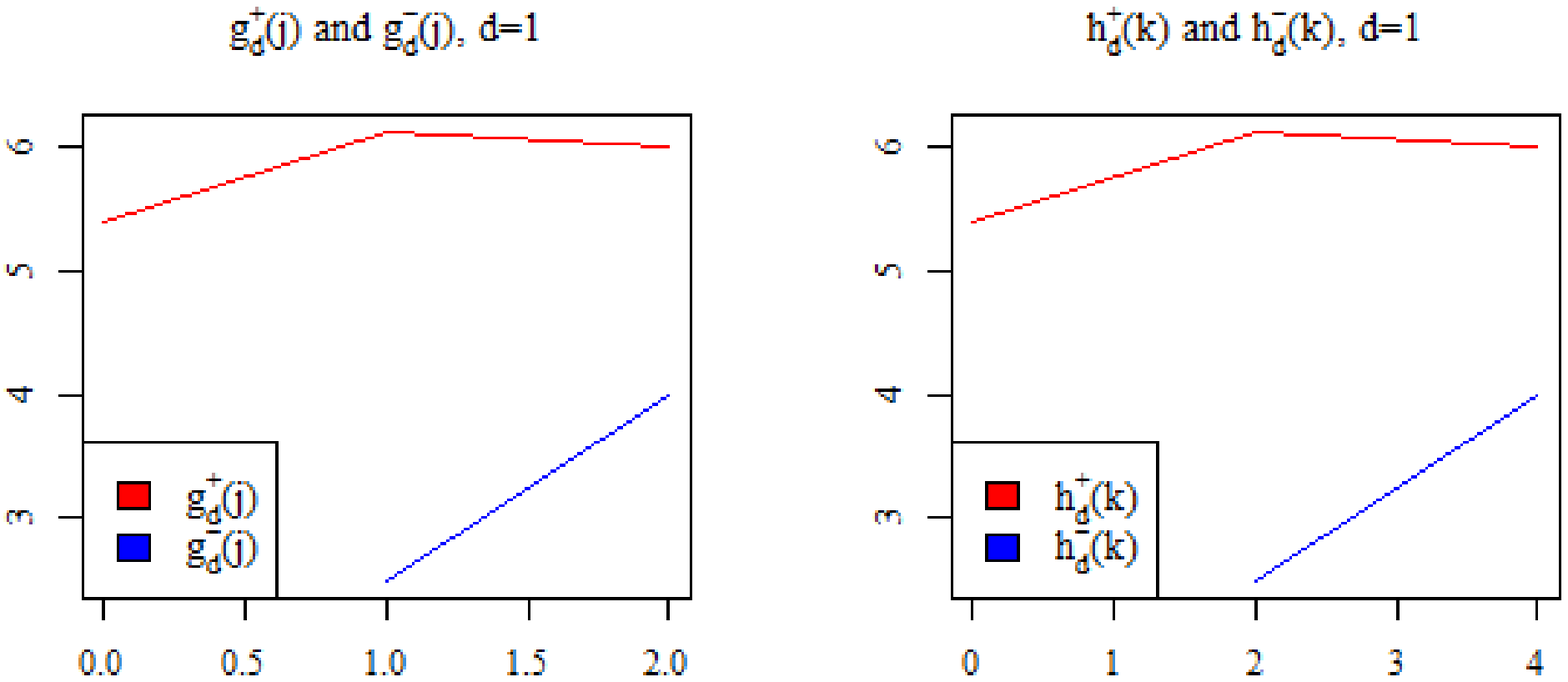}
    \centering
    \caption{Numerical illustration of binarized MTS-MIV 2}
    \label{fig.gh2.exam0921}
\end{figure}

\begin{figure}[ht]
    \includegraphics[width=0.7\textwidth]{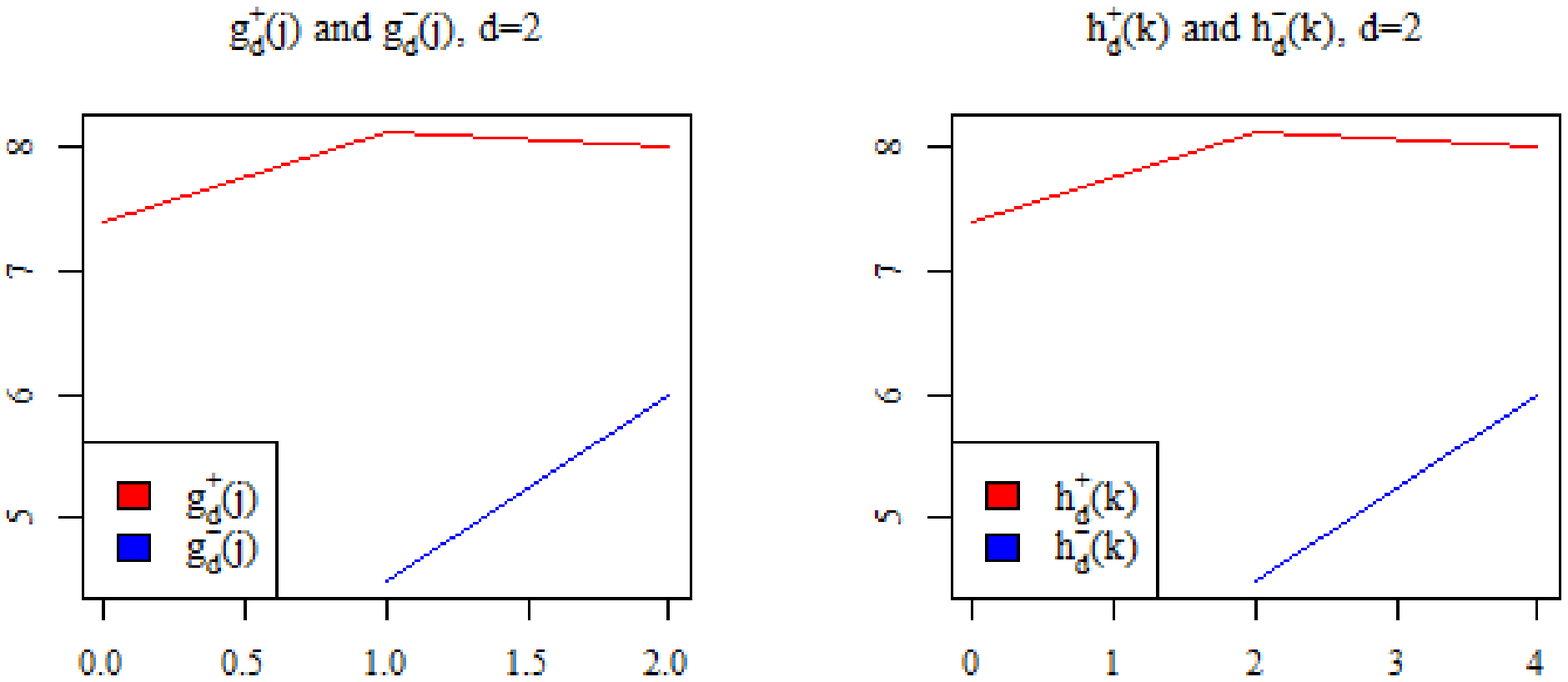}
    \centering
    \caption{Numerical illustration of binarized MTS-MIV 3}
    \label{fig.gh3.exam0921}
\end{figure}

\begin{example}
\label{ex.0921}
\textnormal{
Consider the following data generating process (DGP)}
\begin{eqnarray}
\left\{ \begin{array}{lcl}
    Y &=& 2D + U \\ 
    D &=& 0 \cdot \mathbbm{1}\left\{V \in [0, 1] \right\} + 1 \cdot \mathbbm{1}\left\{V \in (1, \frac{3}{2}] \right\} + 2 \cdot \mathbbm{1}\left\{V \in (\frac{3}{2}, 5] \right\} \\
    Z &=& 2D \\ 
    U &=& 4V \mathbbm{1}\left\{V \in [1, 2] \right\} + V \mathbbm{1}\left\{V \notin [1, 2]\right\}
    \end{array} \right.
\end{eqnarray}
\textnormal{where $V \sim \mathcal U_{[0,5]}$.}

\textnormal{
This example illustrates a case where MTS and MIV fail to hold, but binarized MTS-MIV holds, suggesting that binarized MTS-MIV is a strictly weaker assumption than MTS-MIV (i.e., the converse of Lemma \ref{lem1} does not hold).} 
\textnormal{
In particular, it can be shown that the conditional expectation function $\text{E}\left[Y_d \vert D=\ell\right]$ in the given DGP is not monotone in $\ell$ for each $d=0, 1, 2$, implying that MTS fails. Likewise, the conditional expectation function $\text{E}\left[Y_d \vert Z=z\right]$ is not monotone in $z$ for each $d=0, 1, 2$, which implies that MIV is violated. Figure \ref{fig.ey1.exam0921} illustrates these facts for the case $d=1$.}

\textnormal{
On the other hand, Figures \ref{fig.gh1.exam0921}-\ref{fig.gh3.exam0921} shows that the same DGP satisfies the binarized MTS-MIV restriction (\ref{eq:BMTSMIV}), and accordingly the SDC assumption by Lemma \ref{lem2}, as the function $g_d^+$ is always greater than the function $g_d^-$, and the function $h_d^+$ is always greater than the function $h_d^-$ for all $d$.
}
\end{example}


\end{document}